\documentclass[reprint,amsmath,amssymb,aps,prc,groupedaddress,showpacs,nofootinbib]{revtex4-1}

\usepackage[pdftex]{graphicx}
\usepackage{dcolumn}
\usepackage{bm}
\usepackage{color}

\definecolor{gray}{rgb}{0.8,0.8,0.8}

\newcommand{\be}{\begin{equation}}
\newcommand{\ee}{\end{equation}}
\newcommand{\bea}{\begin{eqnarray}}
\newcommand{\eea}{\end{eqnarray}}
\newcommand{\bfr}{\mbox{\boldmath $r$}}

\newcommand{\bfq}{\mbox{\boldmath $q$}}

\newcommand{\bftau}{\mbox{\boldmath $\tau$}}

\newcommand{\mbss}[1]{_{\mbox{\scriptsize #1}}}
\newcommand{\mbts}[1]{_{\mbox{\tiny #1}}}

\newcommand{\vphu}{\vphantom{*}}
\newcommand{\vphd}{\vphantom{1}}
\newcommand{\vphuu}{\vphantom{(2)}}

\newcommand{\ds}{\displaystyle}

\newcommand{\ve}{\varepsilon}

\newcommand{\vphi}{\varphi}

\begin{document}

\title{The phonon-coupling model for Skyrme forces}

\author{N. Lyutorovich}
\author{V. Tselyaev}
\affiliation{V. A. Fock Institute of Physics, St. Petersburg State University,
             RU-198504 St. Petersburg, Russia}
\author{J. Speth}
\email{J.Speth@fz-juelich.de}
\author{S. Krewald}
\affiliation{Institut f\"ur Kernphysik, Forschungszentrum J\"ulich, D-52425 J\"ulich, Germany}
\author{P.-G. Reinhard}
\affiliation{Institut f\"ur Theoretische Physik II, Universit\"at Erlangen-N\"urnberg,
D-91058 Erlangen, Germany}

\date{\today}

\begin{abstract}
A short review on the self-consistent RPA with in a energy-density functional of the Skyrme type is given. We also present an extension of the RPA where the coupling of phonons to the  single particle states is considered. Within this approach we present numerical results which are compared with data. The self-consistent approach is compared with the Landau-Migdal theory. Here we derive from the self-consistent $ph$ interaction, the Landau-Migdal parameters as well as their density dependence. In the appendix a new derivation of the reduced $ph$ matrix elements is presented.
\end{abstract}


\maketitle

\section{Introduction}

In the present contribution we review the self consistent time
blocking approximation (TBA) and present results obtained within this
method.  The conventional random phase approximation (RPA) allows,
e.g., the calculation of the total transition strength and the mean
energy of giant resonances. If one is interested in the distribution
of the strength and cross sections of various nuclear reactions like
electron scattering and $\alpha$ -scattering it is necessary to
consider in addition to the one-particle--one-hole (1p1h) states more
complex configurations, e.g., two-particle--two-hole states
\cite{Drozdz_1990} or the fragmentation of the single particle
strength due to the coupling to phonons
\cite{Soloviev_1976,Krewald_1977,Tselyaev_1989,Kamerdzhiev_1993}.  As input data for those
approaches one needs single particle energies and wave functions and a
residual particle-hole interaction. In earlier works one started with
a single particle model and defined a residual particle-hole
interaction where both quantities where adjusted to experimental data
\cite{Kamerdzhiev_2004}.  The next stage of
sophistication are self-consistent calculations where one starts with
an effective Hamiltonian or a density functional from which one
obtains by functional derivations the single particle quantities as
well as the residual interaction
\cite{Lyutorovich_2012,Lyutorovich_2015}. The phenomenology enters
here in the Hamiltonian and the density functional in which remain
free parameters which have to be adjusted to experimental data
\cite{Ring1,Bender_2003,Goriely_2002,Nazarevicz1}.

 In the formal part we present the self-consistent RPA equations within an energy-density functional approach and give a short derivation of the TBA. The self-consistent approach is compared with the Landau-Migdal theory. Here we connect the "Landau-Migdal interaction" with the self-consistent $ph$-interaction and calculate the famous Landau-Migdal parameters as well as their density dependence.  

 In all our calculations we started from a density functional of the
Skyrme type where we used various sets of parameters. We found from
our investigations that the inclusion of phonons give very good
results for $^{208}$Pb and fair results for the Ca-isotopes for the
giant isovector dipole resonance (GDR) as well as for the isoscalar
resonances for some of the Skyrme parametrizations. A parameter set
which reproduces all these properties simultaneously is still
missing. In the light nucleus $^{16}$O the GDR is well reproduces but
not the isoscalar resonances. This will be discussed later.

An important point in the calculation of giant resonances is the
treatment of the single particle continuum. In earlier calculations
the continuum was discretized which introduced some
arbitrariness. Most recently we developed a new method for handling
the continuum that allows to consider also the spin-orbit
interaction. Therefore we can now present results of the
self-consistent calculations which include also this part of the
residual interaction.

In Chapter II we introduce the basic formulas of the self-consistent
RPA and TBA.  In Chapter III.A we introduce several different Skyrme
parametrization and discuss their connections to nuclear matter
properties. In Chapter III.B we compare the residual interaction of
the Migdal type with the one which follows from the second derivative
of the Skyrme type density functional. We derive from the residual
Skyrme interaction the spin independent Migdal parameters and their
density behavior.  Finally we present in Chapter IV a large body of
data which are compared with experimental data and in Chapter V we
give a summary.

 In the appendix we discuss in the subsection 1-4 in some detail the Skyrme energy-density functional in general and give detailed formulas for the Skyrme interaction. In subsection 5-9 we derive the reduced matrix elements of the residual $ph$-interaction which include not only the zero-range and velocity dependent parts but also the spin-orbit matrix elements.

\section{The method}
\label{sect2}

\subsection{Formulation of the Hamiltonian}
\label{sec:ham}

RPA is, in principle, a very general method in the context of many-body
theory which emerges from collecting a subset of leading diagrams
(bubble series) within the Green function formalism, for details and
explicit expressions see \cite{Speth_1977}. We will discuss here RPA
in the context of effective interactions, or effective energy-density
functionals respectively (for the expressions in a nuclear context see
\cite{Rei92a,Rei92b}). This will result in somewhat different
notations and aspects. Thus we will introduce here the basic
ingredients as we will employ them in the sequel.

In many-body theory the one-particle Green function is determined by
the mass operator $\Sigma$ which incorporates all information on the
many-body system. Its most general form is
\begin{subequations}
\begin{equation}
  \Sigma
  =
  \Sigma(\mathbf{r} ,\mathbf{p},\epsilon)
\end{equation}
depending on the coordinate $\mathbf{r}$, the momentum $\mathbf{p}$
(non-locality) and the single-particle (s.p.) energy $\epsilon$
(dependence on the spin and isospin variables is tacitly implied).
This $\Sigma$ is a highly non-local one-body operator.  The first
non-local feature is the effective mass which gathers contributions
from $\mathbf{p}$-dependence as well as $\epsilon$-dependence
\cite{Jeukenne_1976}. RPA is a theory for the response function $R$
which is a propagator for a coherent superposition of $1p1h$ states.
It involves, besides $\Sigma$, a residual two-body interaction which
is derived in the many-body framework as
\begin{equation}
  K
  =
  \frac{\delta \Sigma}{\delta G}
\label{eq:Kdef}
\end{equation}
\end{subequations}
where $\delta$ stands for functional derivative and $G$
for the one-body Greens operator. Similar as $\Sigma$, $K$ is a highly
non-local and energy-dependent operators, now even more involved
because it acts on two particles.

Effective interactions as the Landau-Migdal interaction
\cite{Migdal67,LanLif9} or the Skyrme force
\cite{Sky59a,Neg72a,Brink72} are free from energy dependence, but may
carry some momentum dependence of second order in $\mathbf{p}$, for
detailed discussion see section \ref{sec:Skyrme}.  In these cases the
equations-of-motion become much simpler, e.g., as the single-particle
strength is simply one throughout \cite{Frank_2006}.  This is also
indicated by using slightly different notations. The place of
$\Sigma$-operator is now taken by the mean-field Hamiltonian
\begin{subequations}
\begin{equation}
  \hat{h}
  \equiv
  h^{\vphu}_{12}
\end{equation}
which is a one-body operator at most of order $\mathbf{p}^2$. The
numerical indices here and in the following denote the set of the
quantum numbers of some single-particle basis. In the context of the
density functional theory (DFT) with the energy density functional
$E[\rho]$, the mean-field Hamiltonian is deduced from $E[\rho]$ by
functional derivative
\begin{equation}
  h^{\vphu}_{12}
  =
  \frac{\delta E}{\delta\rho^{\vphu}_{21}}\,.
\label{sphdft}
\end{equation}
In practice, RPA described excitation about the stationary ground
state and we deal with the ground-state mean-field Hamiltonian
that is simultaneously diagonal  with $\rho$.
It is convenient to write it in the basis which diagonalizes operators
$h$ and $\rho\,$, yielding
\begin{equation}
  h^{\vphu}_{12} = \varepsilon^{\vphu}_{1}\delta^{\vphu}_{12}\,,
  \qquad
  \rho^{\vphu}_{12} = n^{\vphu}_{1}\delta^{\vphu}_{12}\,,
\label{spbas}
\end{equation}
\end{subequations}
where $n^{\vphu}_{1}=0,1$ is the occupation number.
In what follows the indices $p$ and $h$ will be used to label
the single-particle states of the particles ($n^{\vphu}_{p} = 0$)
and holes ($n^{\vphu}_{h} = 1$) in this basis.

The residual two-body interaction for RPA is derived again from eq.
(\ref{eq:Kdef}), now replacing $\Sigma$ by $\hat{h}$ and $G$ by
$\hat{\rho}$. This yields for the case of DFT eventually
\begin{equation}
  {V}^{\vphu}_{12,34}
  =
  \frac{\delta^2 E[\rho]}
  {\delta\rho^{\vphu}_{21}\,\delta\rho^{\vphu}_{34}}\,,
\label{sccond}
\end{equation}
so the quantities $h$ and $V$ appear to be linked by
Eqs. (\ref{sphdft}) and (\ref{sccond}). Other effective interactions,
as the Landau-Migdal interaction, are modeled directly as residual
two-body interaction at the level of $K$.

\subsection{Self-consistent RPA}

Our approach is based on the version of the response function
formalism developed within the Green function method (see
Ref.~\cite{Speth_1977}).  The aim is to compute a nuclear transition
strength for an observable corresponding to some one-body operator
$Q$. The strength function $S_Q(E)$ is defined in terms of the
response function $R(\omega)$ by
\begin{subequations}
\begin{eqnarray}
  S(E)
  &=&
  -\frac{1}{\pi}\;\mbox{Im}\,\Pi(E+i\Delta)\,,
\label{sfdef}
\\
  \Pi(\omega)
  &=&
  - \langle\,Q\,|\,R(\omega)\,|\,Q\,\rangle\,,
\label{poldef}
\end{eqnarray}
\end{subequations}
where $E$ is the excitation energy, $\Delta$ a smearing parameter
simulating broadening effects beyond RPA, and $\Pi(\omega)$ the
(dynamic) $Q$-polarizability.

The strength functions combines system properties with an observable.
The system property is the response function $R$ which at RPA level is
a solution of the Bethe-Salpeter equation (BSE)
\begin{equation}
  R^{\mbss{RPA}}_{\vphd}(\omega)
  =
  R^{(0)}_{\vphd}(\omega)
  -
   R^{(0)}_{\vphd}(\omega)VR^{\mbss{RPA}}_{\vphd}(\omega)\,,
\label{rfrpa}
\end{equation}
where $R^{(0)}_{\vphd}(\omega)$ is the uncorrelated $1p1h$
propagator and $V$ is the residual two-body interaction
(see section \ref{sec:ham}).
The $1p1h$ propagator $R^{(0)}_{\vphd}(\omega)$ is defined as
\begin{equation}
  R^{(0)}_{\vphd}(\omega)
  =
  -\bigl(\,\omega -
  \Omega^{(0)}_{\vphd}\bigr)^{-1}M^{\mbss{RPA}}_{\vphd}
  \,,
\label{rf0}
\end{equation}
where the matrices $\Omega^{(0)}_{\vphd}$ and $M^{\mbss{RPA}}_{\vphd}$
are defined in the $1p1h$ configuration space. $M^{\mbss{RPA}}_{\vphd}$
is the metric matrix defined (in the diagonal basis) as
\begin{equation}
  M^{\mbss{RPA}}_{12,34}
  =
  \delta^{\vphu}_{13}\,\delta^{\vphu}_{24}\,
  \left[n^{\vphu}_{2}-n^{\vphu}_{1}\right]
  \,.
\label{mrpa}
\end{equation}
The matrix $\Omega^{(0)}_{\vphd}$ comprises the one-body Hamiltonian
acting separately on particle and hole in the form
\begin{equation}
  \Omega^{(0)}_{12,34}
  =
  \delta^{\vphu}_{13}\,\delta^{\vphu}_{24}\,
  \left[\varepsilon^{\vphu}_{1}-\varepsilon^{\vphu}_{2}\right]
  \,.
\label{omrpa}
\end{equation}

The propagator $R^{\mbss{RPA}}_{\vphd}(\omega)$, being a matrix in
$1p1h$ space, is a rather bulky object. For practical calculations, it
is more convenient to express it in terms of RPA amplitudes $z_{12}^n$
by virtue of the spectral representation
\begin{equation}
  R^{\mbss{RPA}}_{1234}(\omega)
  = -\sum_nz_{12}^n
  \frac{\mbox{sgn}(\omega_n)}
  {\omega-\omega_n}(z_{34}^n)^*
\end{equation}
where $n$ labels the RPA eigenmodes and $\omega_n$ is the eigenfrequency.
Inserting that into Eq.~(\ref{rfrpa}) and filtering the pole at
$\omega=\omega_n$ yields the familiar RPA equations
\begin{equation}
 \sum_{34}\left(\Omega^{(0)}_{12,34}
    + \sum_{56} M^{\mbss{RPA}}_{12,56}\,{V}^{\vphu}_{56,34}\right)
    \,z^{n}_{34}
=
  \omega^{\vphu}_n\,z^{n}_{12}\,,
\label{rpaze}
\end{equation}
where the transition amplitudes ${z}^{n}$ are normalized
by the condition
\begin{equation}
 \sum_{12,34}({z}^{n}_{12})^*\,M^{\mbss{RPA}}_{1234}\,{z}^{n'}_{34}
 =
 \mbox{sgn}(\omega^{\vphu}_{n})\,\delta^{\vphu}_{n,\,n'}.
\label{zmz}
\end{equation}
These equations determine the set of eigenstates $n$ with amplitudes
$z^{n}_{12}$ and frequencies $\omega_n$.

\subsection{Phonon coupling model}

The second model is the quasiparticle-phonon coupling model within
the time-blocking approximation (TBA)
\cite{Tselyaev_1989, Kamerdzhiev_1997, Kamerdzhiev_2004, Tselyaev_2007}
(without ground state correlations beyond the RPA included in
\cite{Tselyaev_1989, Kamerdzhiev_1997, Kamerdzhiev_2004, Tselyaev_2007} and without pairing correlations
included in \cite{Tselyaev_2007}).
This model, which in the following will be referred to as the TBA,
is an extension of the RPA including $1p1h\otimes$phonon
configurations in addition to the $1p1h$ configurations incorporated
in the conventional RPA.
The BSE for the response function in the TBA is
\begin{eqnarray}
  R^{\mbss{TBA}}_{\vphd}(\omega)
  &=&
  R^{(0)}_{\vphd}(\omega) \nonumber \\
  &&-
  R^{(0)}_{\vphd}(\omega)(V\!+\!\tilde{W}(\omega))
  R^{\mbss{TBA}}_{\vphd}(\omega) \,,
\label{rftba}\\
  \tilde{W}(\omega)
  &=& W(\omega)-{W}(0)\,,
\label{Wsubtract}
\end{eqnarray}
where the induced interaction $\tilde{W}(\omega)$
serves to include contributions of $1p1h\otimes$phonon configurations.
The matrix ${W}(\omega)$ in Eq. (\ref{Wsubtract})
is defined in the $1p1h$ subspace and can be represented in the form
\begin{subequations}
\begin{equation}
  {W}^{\vphu}_{12,34}(\omega)
  =
  \sum_{c,\;\sigma}\,\frac{\sigma\,{F}^{c(\sigma)}_{12}{F}^{c(\sigma)*}_{34}}
                   {\omega - \sigma\,\Omega^{\vphu}_{c}}
  \,,
\label{wdef}
\end{equation}
where $\sigma = \pm 1$, $\,c = \{p',h',n\}$ is an index of the subspace
of $1p1h\otimes$phonon configurations, $n$ is the phonon's index,
\begin{eqnarray}
  &
  \Omega^{\vphu}_{c}
  =
  \ve^{\vphu}_{p'} - \ve^{\vphu}_{h'} + \omega^{\vphu}_{n}
  \,,\quad
  \omega^{\vphu}_{n}>0
  \,,
\label{omcdef}
\\&
  {F}^{c(-)}_{12}={F}^{c(+)*}_{21}
  \,,\qquad
  {F}^{c(-)}_{ph}={F}^{c(+)}_{hp}=0\,,
\label{fcrel}
\\&
  {F}^{c(+)}_{ph}
  =
  \delta^{\vphu}_{pp'}\,g^{n}_{h'h} -
  \delta^{\vphu}_{h'h}\,g^{n}_{pp'}
  \;,
\label{fcdef}
\end{eqnarray}
and $g^{n}_{12}$ is an amplitude of the quasiparticle-phonon
interaction.
These $g$ amplitudes (along with the phonon's energies
$\omega^{\vphu}_{n}$) are determined by the positive frequency
solutions of the RPA equations and the emerging $z$ amplitudes as
\begin{equation}
g^{n}_{12} = \sum_{34} {V}^{\vphu}_{12,34}\,z^{n}_{34}\,.
\label{gndef}
\end{equation}
\end{subequations}

In our DFT-based approach the energy density functional $E[\rho]$ in
Eqs. (\ref{sphdft}) and (\ref{sccond}) is the functional of the Skyrme
type with model parameters adjusted to reproduce nuclear
ground state properties with high quality. In this case $E[\rho]$
already effectively contains a part of the contributions of those
$1p1h\otimes$phonon configurations which are explicitly included in
the TBA. Therefore, in a theory going beyond the RPA, the problem of
double counting arises.  To avoid this problem in the TBA, we use the
subtraction method.  It consists in the replacement of the amplitude
${W}(\omega)$ by the quantity $\tilde{W}(\omega)={W}(\omega)-{W}(0)$
as it is written in Eq.~(\ref{rftba}). In Ref.~\cite{Tselyaev_2013} it
was shown that, in addition to the elimination of double counting, the
subtraction method ensures stability of solutions of the TBA
eigenvalue equations.

\section{Effective interactions}
\label{seq:effint}

\subsection{Basics on the Skyrme functional and related parameters}
\label{sec:Skyrme}

From the variety of self-consistent nuclear mean-field models
\cite{Bender_2003}, we consider here a non-relativistic branch, the
widely used and very successful Skyrme-Hartree-Fock (SHF) functional.
A detailed description of the functional is given in the appendix
 and more background information can be found in the
reviews \cite{Bender_2003,Stone_2007,Erler_2011}. We summarize here
the essential features. The functional depends on a couple of local
densities and currents (density, gradient of density, kinetic-energy
density, spin-orbit density, current, spin density, kinetic
spin-density). For the description of ground state properties and
natural-parity excitations, there remain typically 13--14 free model
parameters.  They are usually determined
by a fit to a large and representative set of experimental data on
bulk properties of the nuclear ground state, for recent adjustments
see \cite{Goriely_2002,Kluepfel_2009,Kortelainen_2010}. The parameters
thus found are considered to be universal parameters as they apply to
all nuclei throughout the nuclear landscape and to astro-physical
matter (e.g. neutron stars).

The force parameters $C_T^\mathrm{(typ)}$, although necessary for
communicating the model and coding, are not very intuitive.  The most
important part of the functional can be characterized by nuclear
matter properties (NMP), i.e.  equilibrium properties of homogeneous,
symmetric nuclear matter, for which we have some intuition from the
liquid-drop model (LDM) \cite{Myers_1977}.  A detailed definition of
the NMP is given in  appendix A4.  Of particular interest
for resonance excitations are the NMP which are related to response to
perturbations: incompressibility $K$ (isoscalar static), effective
mass $m^*/m$ (isoscalar dynamic), symmetry energy $a_\mathrm{sym}$
(isovector static), TRK sum rule enhancement $\kappa_\mathrm{TRK}$
(isovector dynamic).  It turns out that a fit to ground state
properties fixes some of the NMP very well (equilibrium density and
binding energy, to some extend also incompressibility) while others
are left with an appreciable leeway (particularly the isovector
properties symmetry energy and sum-rule enhancement).  This calls for
careful evaluation of the predictive value of SHF calculations. There
is a great manifold of strategies to explore the uncertainties in
predictions, for recent discussion see, e.g.,
\cite{Dob14a,Erl14b,Rei15d}.

One such strategy is a systematic variation of properties of a
functional in the vicinity of the optimal fit. And it is obvious that
one should vary the most important agents which are for the SHF
functional the NMP. In that spirit, the paper \cite{Kluepfel_2009}
provides a series SHF parametrizations with systematically varied
NMP. We use these sets here to explore the sensitivity, or robustness,
of the phonon coupling under variations of the functional in
reasonable ranges (i.e. in the vicinity of the optimum).
\begin{table}
\centering
\begin{tabular}{l|cccc}
  & $K$ [MeV]& $m^*/m$ & $a_\mathrm{sym}$ [MeV]& $\kappa_\mathrm{TRK}$ \\
\hline
SV-bas &  234 &  0.90  &  30  &  0.4 \\
SV-kap00 & 234 &  0.90  &  30  &  0.0 \\
SV-mas07 & 234 &  0.70  &  30  &  0.4 \\
SV-sym34 & 234 &  0.90  &  34  &  0.4 \\
SV-K218 & 218 &  0.90  &  30  &  0.4 \\
SV-m64k6 & 241 & 0.64  & 27 & 0.6\\
SV-m56k6 &  255 &   0.56 &   27 &   0.6\\
\hline
\end{tabular}
\caption{\label{tab:NMP} NMP for the Skyrme paramterizations used in
  this study: incompressibility $K$, isoscalar effective mass $m^*/m$,
  symmetry energy $a_\mathrm{sym}$, Thomas-Reiche-Kuhn sum rule
  enhancement $\kappa_\mathrm{TRK}$. The first five parametrizations
  stem from \cite{Kluepfel_2009}, the last two from
  \cite{Lyutorovich_2012}. For the definition of the NMP, see appendix
  \ref{app:SHF}.}

\end{table}
Table \ref{tab:NMP} lists the selection of parametrizations and their
NMP. SV-bas is the base point of the variation of forces. Its NMP are
chosen such that dipole polarizability and the three most important
giant resonances (GMR, GDR, and GQR) in $^{208}$Pb are well reproduced
by Skyrme-RPA calculations. Each one of the next four parametrizations
vary exactly one NMP while keeping the other three at the SV-bas
value. These 1+4 parametrizations allow to explore the effect of each
NMP separately.  It was figured out in \cite{Kluepfel_2009} that there
is a unique relation between each one of the four NMP and one
resonance in $^{208}$Pb: $K$ affects mainly  the GMR, $m^*/m$ affects mainly
the GQR, $\kappa_\mathrm{TRK}$ affects mainly the GDR, and
$a_\mathrm{sym}$ is uniquely linked to the dipole polarizability
\cite{Nazarewicz_2013}.
Finally, the last two parametrizations in table
\ref{tab:NMP} were developed in \cite{Lyutorovich_2012} with the goal
to describe, within TBA, at the same time the GDR in $^{16}$O and
$^{208}$Pb. This required to push up the RPA peak energy which was
achieved by low $a_\mathrm{sym}$ in combination with high
$\kappa_\mathrm{TRK}$. To avoid unphysical spectral distributions for
the GDR, a very low $m^*/m$ was used.

\subsection{From Landau-Migdal theory to the SHF residual interaction}
\label{sec:LM}

The SHF theory provides reliable nuclear ground states and with it the
first ingredients for the RPA/TBA equations, namely
s.p. wavefunctions, mean-field Hamiltonian $\hat{h}$ and
corresponding s.p. energies $\varepsilon_i$. The crucial piece, going
beyond ground state properties, is the effective residual interaction.
It can be derived fully self-consistency from the SHF functional with
Eq. (\ref{sccond}). The details of its evaluation for spherical nuclei
are given in appendix A5-A9. In these sections, we will briefly
review the development of effective residual interactions which
started out with the Landau-Migdal interaction in the context of Fermi
liquid theory \cite{Migdal67} and compare it with the actual form
delivered by the SHF functional, for a more extensive discussion see
\cite{NPA928}.

The theory of Fermi liquids deals with homogeneous matter where
momentum space provides the most natural representation. We will thus
discuss in the following the residual interaction in momentum space.
In general, the effective interaction kernel is an involved four-point
function, depending on three momenta:
$\mathbf{p}$ and $\mathbf{p'}$ as the momenta of the in-coming and
out-going hole states and $\mathbf{q}$ as the transferred momentum.
\begin{figure}[htbp]
\begin{center}
\includegraphics[width=0.7\linewidth]{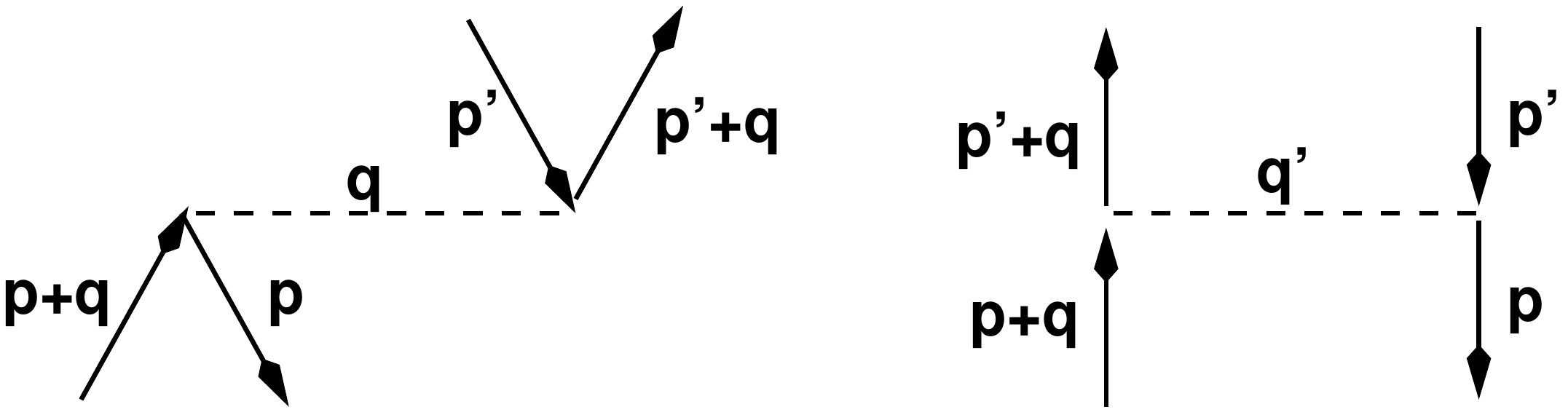}
\end{center}
\caption{\label{fig:basicV}
Graphical representation of a local interaction in $ph$ space
with direct (left) and exchange term (right). Particle and
hole states are represented by full lines with up- and down arrows.
The dashed horizontal line stands for the interaction.
The three relevant momenta $\mathbf{p}$,  $\mathbf{p}'$, and
$\mathbf{q}$, are indicated. The exchange term transfers the
momentum  $\mathbf{q}'=\mathbf{p}'-\mathbf{p}$.
}
\end{figure}
Figure \ref{fig:basicV} illustrates these momenta for the case of a
local interaction which is sufficient for our purposes because the
Landau-Migdal as well as the Skyrme interaction are both local.  All
$ph$ pairs carry net momentum $\mathbf{q}$. They differ by the other
momentum $\mathbf{p}$, or $\mathbf{p}'$ respectively.  In Landau
approximation, one considers the residual interaction $F^{ph}$ at the
Fermi surface (we use here the symbol $F$ to distinguish Landau-Midgal
approach from $V$ in SHF). Thus one can approximate it as a local
contact (zero-range) interaction and $F^{ph}$ depends not on the
$\mathbf{q}$ any more, only on the angle between the momenta
$\mathbf{p}$ and $\mathbf{p'}$ before and after the collision; its
spin-independent part reads
\begin{eqnarray}\label{eq:3a}
  F^{ph}\left( \frac{ \mathbf{p}\cdot \mathbf{p'}} {p^{2}_F} \right)
  &=&
  C_0\sum_{l=0}^{\infty}
  \left[f_l+f^{\prime}_l\hat{\mathbf{\tau_1}}\cdot\hat{\mathbf{\tau_2}}
  \right]
      P_l\left( \frac{ \mathbf{p}\cdot \mathbf{p'}} {p^{2}_F} \right)
\\
  C_0 &=& \frac{\pi^2 \hbar^3}{2 m^* p_{\rm F} }
  \quad,\quad
  p_\mathrm{F}=\hbar\left(\frac{3\pi^2\rho_0}{2}\right)^{1/3}
\end{eqnarray}
where $P_l(x)$ is the Legendre polynomial of order $l$ and $p_F$ is
the Fermi momentum. There are, in fact, four terms containing
different combinations of spin and isospin operators covering the typical
nuclear interaction channels \cite{Rin80aB}. We consider here only the
two terms which are relevant for modes with natural parity as, e.g.,
the giant resonance.  By virtue of the Landau quasi-particle concept,
the whole information content of the two-body interaction in matter
shrinks to a few model constants, the much celebrated Landau-Migdal
(dimensionless) parameters $f_l$.  The scaling factor $C_0$ is
proportional to the density of states at the Fermi surface.  A typical
value is $C_0=150\,\mathrm{MeV}\,\mathrm{fm}^3$ which is the standard
choice in phenomenological shell models where the effective mass is
$m^*/m=1$. The importance of the $f_l$ shrinks with increasing
$l$. Usually, only $l=0$ and 1 are taken into account.
The Fourier transforms of the terms with $l=0$ and $l=1$ yield
$\delta$-functions in coordinate space and derivatives thereof, a form
which resembles very much the Skyrme force as given in
Eq. (\ref{eq:SHFforce}). This already indicates that there is close
relation between Landau-Migdal theory and SHF.

In the \emph{Theory of Fermi Liquids} the Landau parameters
are constants. Migdal introduced in his \emph{Theory of Finite Fermi
  Systems} density dependent parameters $f_l(\rho)$ in order to
correct for the finite size of the nuclei.
The density dependent Landau-Migdal parameters are parametrized as
\cite{Migdal67}:
\begin{equation}\label{eq:22}
  f(\rho)
  =
  f^\mathrm{(ex)} + (f^\mathrm{(in)}-f^\mathrm{(ex)})\frac{\rho_0(r)}{\rho_0(0)}
\end{equation}
where $f^\mathrm{(ex)}$ stands for the exterior region of the nucleus
and $f^\mathrm{(in)}$ for the interior.
The Landau-Migdal interaction describes one part of the RPA scheme,
the residual interaction. In Landau-Migdal theory, the ground-state
input (s.p.-wavefunctions, s.p. energies) is taken 
from an empirical single-particle model which reproduces experimental
s.p. energies as good as possible.  The Landau-Migdal
parameters
are free parameters of the model and tuned to generally accepted values of NMP,
incompressibility, symmetry energy and effective masses, for details
see \cite{Kamerdzhiev_2004}.

SHF goes one step further in that it describes ground states as well
as excitation properties with one and the same energy functional.  The
residual interaction for RPA calculations is derived with
Eq. (\ref{sccond}), for details see appendix A5-A9. The terms
for natural-parity modes (no spin-spin interaction) eventually read in
momentum space
\begin{subequations}
\begin{eqnarray}
 F^{ph}_\mathrm{Sk}(\mathbf{p},\mathbf{p}',\mathbf{q})
 &=&
%
a^{\vphuu}_{00} + a^{\vphuu}_{01}(\bftau\bftau')
+ a^{\vphuu}_{z}(\tau^{\vphuu}_{z} + \tau^{\prime\vphuu}_{z})
\nonumber\\
 &&
 +\bigl[b^{(-)}_{00} +b^{(-)}_{01}(\bftau\bftau')\bigr]
 \mathbf{q}^2
\nonumber\\
  &&
   \!\!\!+\bigl[b^{(+)}_{00} +b^{(+)}_{01}(\bftau\bftau')\bigr]
   (\mathbf{p}\!-\!\mathbf{p}')^2\;,
\label{eq:kinF}
\\
  a^{\vphuu}_{0T} 
  &=& 
  \frac{1}{4}\sum_{\tau}\bigl[F^{\,0}_{\tau,\tau} +
             (-1)^T F^{\,0}_{\tau,-\tau}\bigr],
\nonumber
\\
  a^{\vphuu}_{z} 
  &=& 
  \frac{1}{4}\sum_{\tau}(\tau^{\vphuu}_{z})^{\vphuu}_{\tau,\tau}F^{\,0}_{\tau,\tau},
\end{eqnarray}
\end{subequations}
where $F^{\,0}_{\tau,\tau}$, $F^{\,0}_{\tau,-\tau}$ ($\tau = n,p$) and
the $b$ parameters are given in appendix \ref{append1}.  This is the
residual interaction as it must be taken into account in a consistent
Skyrme-RPA calculation of modes with natural parity.  It is
interesting to check the expression in Landau approximation which
reads in the limit of nuclear matter
\be
\bfq = 0\quad,\quad \mathbf{p}^2 = \mathbf{p}'^2 = k_F^2\,.
\label{qpppnm}
\ee
yielding
\be
  \bfq'^2
  =
  2k_F^2\bigl[\,1 - P_1(\cos\theta)\,\bigr]
  \quad,\quad
  \cos\theta
  =
  \frac{\mathbf{p}\cdot\mathbf{p}'}{k_F^2}
  \quad.
\label{qp2nm}
\ee
This erase the direct term $\propto\mathbf{q}^2$ leaving for
the momentum dependent part
\begin{eqnarray}
 F^{ph}_\mathrm{Lan,grad}
 &=&
   2k_F^2\bigl[b^{(+)}_{00}+b^{(+)}_{01}(\bftau\bftau')\bigr]
\nonumber\\
  &&
   -2k_F^2\bigl[b^{(+)}_{00}+b^{(+)}_{01}(\bftau\bftau')\bigr]
    P_1(\cos\theta)\,
   \quad.
\label{eq:kinFLan}
\end{eqnarray}
Note that the velocity dependent exchange terms contribute to the
leading order of $F^{ph}$($f_0$, $f'_0$) as well as to the next to
leading order ($f_1$, $f'_1$). Comparing Eq. (\ref{eq:kinFLan}) with
the full form (\ref{eq:kinF}), we see that, in spite of much
similarity, the Landau-Migdal approximation modifies the residual
interaction in detail. The effect of momentum dependence is formally
obvious from Eq. (\ref{eq:kinFLan}).
\begin{figure}
\centerline{\includegraphics[width=\linewidth]{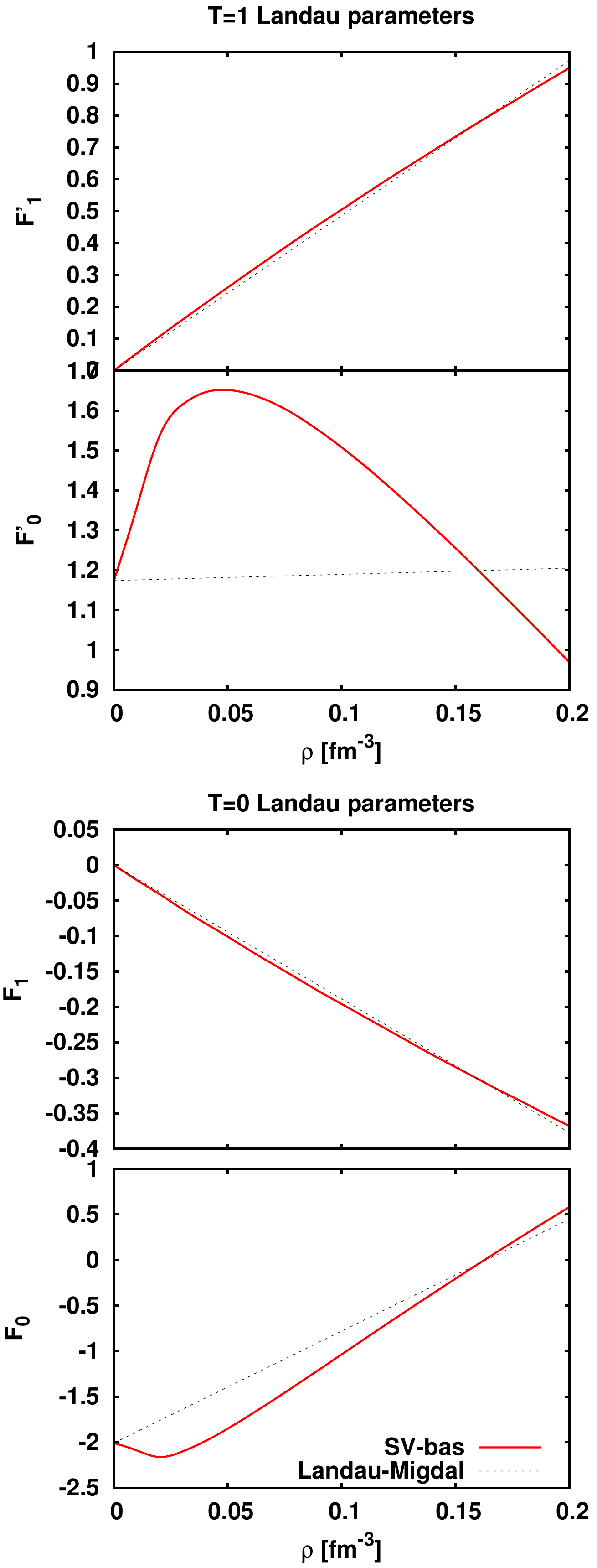}}
\caption{\label{fig:lanpar-uncert1} Dimensionless Landau-Migdal
  parameters for the Skyrme
  parametrization SV-bas \cite{Kluepfel_2009} together with the linear
  approximation according to the Landau-Migdal ansatz (\ref{eq:22}).}
\end{figure}
The effect of density-dependence is illustrated in figure
\ref{fig:lanpar-uncert1} and shows the dimensionless Landau-Migdal
parameters $f_0(\rho)$ and $f^\prime_0(\rho)$ for the SHF
parametrization SV-bas and compares it with the standard linear
Landau-Migdal form (\ref{eq:22}).  The $f_0$ and $f'_0$ parameters
look, at first glance, much different from the linear trend the
Landau-Migdal ansatz.  However, one has to remind that nuclear
resonance excitations do not explore the whole range of densities. The
dynamically most relevant region lies between volume density and
density at the surface $\rho\approx 0.1-0.16$ fm$^{-3}$ and here, the
differences are not so dramatic. The results from an RPA calculation
in Landau-Migdal approximation are basically fine
\cite{Lyutorovich_2008,Avdeenkov_2009}. But for a detailed
description, one should use the full SHF residual interaction.
This has to include also a correct residual interaction from
the other terms in the functional, particularly concerning the Coulomb
interaction \cite{Sil06}.
The density dependence of the $l=1$ parameters, f$_1$ and
f$^\prime_1$, causes no problem as it is also linear in SHF which is
obvious from figure \ref{fig:lanpar-uncert1}.

\section{Results}
\label{sec:results}

From the huge variety of possible results, we concentrate on the three
most important giant resonances: the isoscalar giant monopole
resonance (GMR), the isoscalar giant quadrupole resonance (GQR), and
the isovector giant dipole resonance (GDR). To avoid interference with
pairing effects, we confine the study to the doubly magic nuclei
$^{16}$O, $^{40}$Ca, $^{48}$Ca, and $^{208}$Pb.

\subsection{Details of the calculation scheme}
\label{sec:calc}

Eqs. (\ref{rfrpa}) and (\ref{rftba}) for the response functions in our
approach are solved in a discrete basis. This basis is defined as a
set of solutions of the Schr\"odinger equation with box boundary
conditions.  Both equations are solved in the same large configuration
space. The RPA solutions with sufficiently strong $B(EL)$ values are
taken as the phonons for the TBA calculation. We check stability of
the results with phonon space and show here only the converged
results. A detailed discussion of stability with respect to size of
RPA space, phonon space, and $B(EL)$ cutoff will be postponed to a
forthcoming publications.
Actually, we use for phonon coupling all RPA modes which exhaust more
than 20\% of the total $B(EL)$ strength.
 For final TBA and RPA results we switch to a
description of the nucleon continuum by using the continuum
representation for the free $1p1h$ response propagator $R^{(0)}$ in
eqs. (\ref{rfrpa}) and (\ref{rftba}), details will be presented in a
separate publication.
It is to be noted that continuum effects are marginal for $^{208}$Pb,
but play a significant role for the lighter nuclei in the survey.
 The residual interaction $V$ in
Eqs.(\ref{rfrpa}) and (\ref{rftba}) is derived fully self-consistently
from the SHF functional according to Eq.~(\ref{sccond}).  In the case
of the energy density functional $E[\rho]$ built on the Skyrme forces,
the amplitude $V$ determined by Eq.~(\ref{sccond}) contains the
zero-range (velocity-independent), and velocity-dependent parts, and
the Coulomb interaction. Explicit formulas for all these terms of V are given in Appendices A5-A9. 

We will consider only the doubly-magic nuclei. They have closed shells
and pairing is not important.  The box sizes in the RPA and TBA
calculations are 15 fm for $^{16}$O, $^{40,48}$Ca and 18 fm for
$^{208}$ Pb.  The cutoff for the $1p1h$ space is 100 MeV for all
nuclei (see our discussion in the next two sections).
The new aspects in the present calculations as compared to earlier
presentation are:\\
1) full residual interaction from the SHF functional,\\
2) continuum effects, and\\
3) subtraction method (\ref{Wsubtract}) in TBA.\\
Point 1, the fully residual interaction, assures consistency of the
calculations. Point 2, the particle continuum, serves to model
correctly the escape with in the spectrum. Point 3, the subtraction
of the static contribution from the $1p1h$-phonon-interaction
eliminates the double counting, resolves the stability problem of TBA,
and reinstates the Thouless theorem, which otherwise does not hold for
extended versions of the RPA.

\subsection{The impact of the subtraction scheme}
\label{sec:subtr}

\begin{figure}
\centerline{\includegraphics[width=\linewidth]{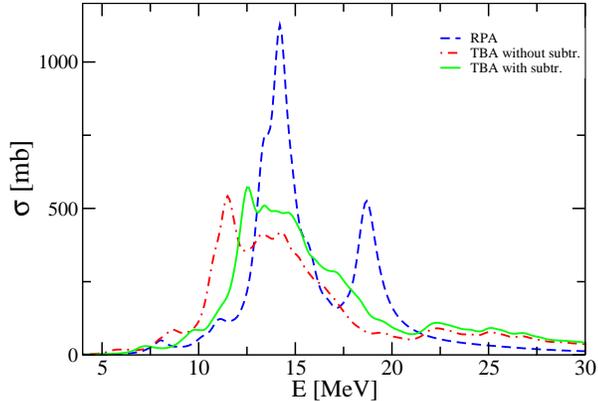}}
\caption{\label{fig:208Pb_photo_m64k6_subtr_effect} Photo-absorption
  strength in $^{208}$Pb calculated with the parameter set
  SV-m64k6. Compared are RPA, plain TBA, and TBA with subtraction
  scheme (\ref{Wsubtract}).  Details of the calculations are the same
  as described in section \ref{sec:calc} with spectral smoothing
  parameter $\Delta = 400$ keV.}
\end{figure}
First, we have a look at the effect of the subtraction scheme
(\ref{Wsubtract}) for the residual interaction in TBA.  Figure
\ref{fig:208Pb_photo_m64k6_subtr_effect} demonstrates that for the
photo-absorption strength in $^{208}$Pb. The RPA spectrum looks
already rather smooth due to the high density of $1p1h$ states,
continuum treatment, and folding with $\Delta = 400$ keV. But there is
a pronounced secondary peak at higher energy around 18 MeV which is
not found in experimental data (see figure
\ref{fig:strength-all-208Pb}). The coupling to complex configurations
in TBA smooths the unnatural high-energy peak and turns it to a long
high-energy tail in the spectrum which is also found in the
experimental spectra. This means that TBA is correctly describing
collisional broadening and both version of TBA do that in similar
manner.  Besides broadening, TBA induces also an energy shift in the
spectra, usually a down shift as seen here. This shift is much reduced
by the subtraction scheme and that is a desirable effect. An example
for this is the isoscalar quadrupole channel in $^{208}$Pb. There is a
strong low lying $2_1^+$ mode around 4 MeV and one expects that it is
robust against complex configurations because the phase space or
collisional effects is too small at this low energy. It turns out that
this mode is heavily down shifted for plain TBA but almost inert, as
it should be, for TBA with subtraction scheme. Altogether, we see that
the subtraction of the zero-frequency interaction $W(0)$ is a crucial
ingredient in TBA.

\subsection{Strength distributions}

\begin{figure*}
\centerline{\includegraphics[width=\linewidth]{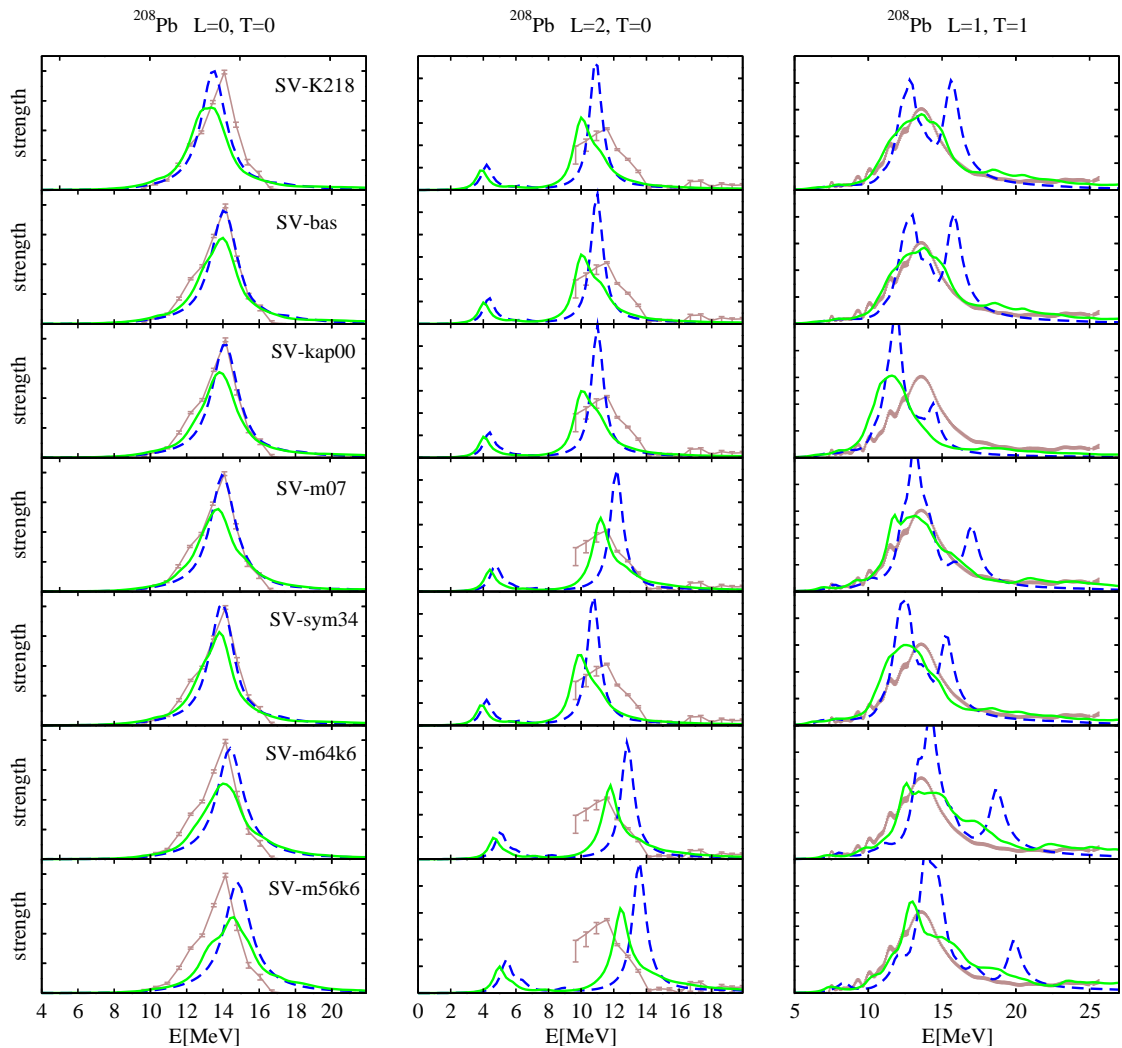}
}
\caption{\label{fig:strength-all-208Pb} Detailed spectral strength
  distributions for $^{208}$Pb and the three modes under
  consideration: isoscalar monopole (left panels), isoscalar
  quadrupole (middle panels), and isovector dipole (right panels).
  Photo-absorption strength is shown in case of the dipole mode,
  multipole strength else-wise.  Compared are results from RPA (blue
  dashed) and TBA (green) with experimental strengths (brown with
  errorbars). Results are obtained with the seven Skyrme
  parametrizations presented in section \ref{sec:Skyrme}.
  Experimental data are from \cite{Belyaev_1995} for the GDR and
  \cite{Youngblood_2004} for GMR and GQR.}
\end{figure*}
In Fig.~\ref{fig:strength-all-208Pb} the theoretical cross sections of
GMR, GQR and GDR are compared with the experimental ones for
$^{208}$Pb.  The theoretical results are calculated with all seven
Skyrme parameter sets which we presented in Table~\ref{tab:NMP} of
section \ref{sec:Skyrme}.  We first discuss the GMR (left column)
which is closely connected with the incompressibility $K$.  The peak
position is clearly related to $K$. Low $K$ (upper left panel) shifts
the peak to lower energy while high $K$ shift it up. All
parametrizations with $K=234$ MeV (left column, panels 2--5 from
above) produce the GMR at the same and correct place although they
differ in other NMP. The GQR is shown in the middle column.  It
confirms what had been found earlier \cite{Bra85aR}, namely that the
GQR depends sensitively on the effective mass $m^*/m$ with the peak
position going up with smaller $m^*/m$. RPA fits best with $m^*/m=0.9$
(panels 1--3 and 5 from above) but misses the high-energy tail. TBA
provides best results with $m^*/m=0.7$ and produces properly the upper
tail of the spectral distribution.  The GDR is shown on the right
column. Most prominent is the unphysical high-energy peak which shows
up for all parametrizations and the welcome feature that TBA removes
it consistently, as was discussed in section \ref{sec:subtr}.  What
trends is concerned, we see for the GDR the strongest impact coming
from $\kappa_\mathrm{TRK}$, see the deviation for SV-kap00 (panel 3
from above). The situation is mixed for SV-m64k6 and SV-m56k6 because
more than one NMP was varied (see table \ref{tab:NMP}).

\begin{figure*}
\centerline{\includegraphics[width=\linewidth]{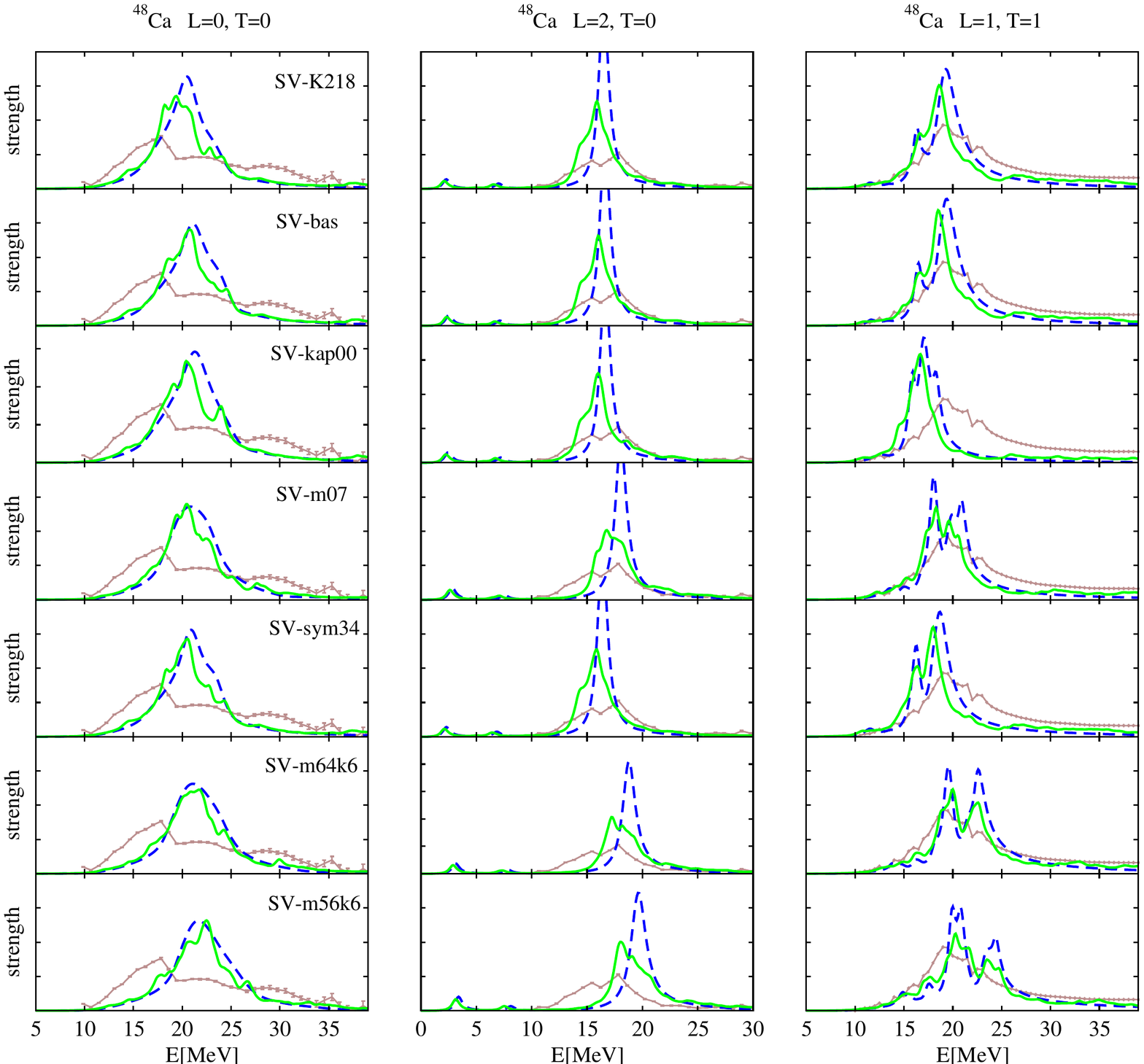}
}
\caption{\label{fig:strength-all-48Ca} As
  Fig.~\ref{fig:strength-all-208Pb}, but for $^{48}$Ca. Experimental
  data are taken from \cite{Erokhova_2003} for the GDR and
  \cite{Anders_2013} for GMR and GQR.}
\end{figure*}
Fig.~\ref{fig:strength-all-48Ca} shows photo-absorption cross section
(isovector dipole channel) and multipole strength distributions
(isoscalar channels) for $^{48}$Ca. The general trends are much the
same as for $^{208}$Pb.  New is that in lighter nuclei spectra become
broader due to more spectral fragmentation.  A closer inspection of
the nucleus $^{48}$Ca may help to assess the importance of the phonon
degree of freedom in theoretical approaches to the spectral strength
distributions. We first look at the GDR strength (right column). The
experimental GDR strength is centered at 20.2 MeV and shows large and
smooth low- and high-energy tails, see
Fig. \ref{fig:strength-all-48Ca}. The mean-field approximation(RPA)
produces a fragmentation of the strength into two major peaks and
overestimates the height of the peaks by approximately a factor 2. The
detailed shape of the strength distribution is sensitive to the chosen
parametrization of the effective interaction. Using a small effective
mass, significant RPA strength is produced at a high energy of 25
MeV(SV-m56K6).  The phonon coupling in TBA mainly reduces the strength
in the vicinity of the two RPA peaks and re-distributes it.  The
effect of TBA is most pronounced for the interaction SV-m56K6 where
the final TBA strength comes closest to the experimental data.
The effect of peak reduction and broadening is more pronounced for the
GQR strength (middle column) where RPA shows a too narrow resonance
that dramatically overestimates the experimental strength of the
peak. The phonon coupling reduces the strength of the RPA peak by a
factor 2. Even this does not yet suffice to obtain a quantitative
description of the experimental quadrupole data. Somehow, collisional
broadening is still underestimated.
This underestimation is even more dramatic for the GMR (left column)
where phonon coupling in TBA makes only a minor modification of the RPA
result.  From a formal point of view, this is plausible because the
L=0 channel limits the possible phonon couplings.  Experimental data,
however, are significantly broadened, more than for the other modes.
It seems that TBA is not properly accounting for the crucial
broadening mechanisms in $^{48}$Ca (and other light nuclei).  This
leaves an open problem for future research.

The main result from the large collection of strengths shown in this
section is that the modifications of the RPA results brought in by TBA
(broadening and shift) are for a given channel and nucleus are
practically the same for all parametrizations. We will see this also
from the compact analysis in the next section.

\subsection{Trends in terms of peak energies}

After looking at strength functions in detail, we want to summarize
here the net effect of TBA in terms of one key number. To that end, we
define a resonance peak energy by averaging the strength in a window
around the resonance.  The peak energy was defined as the energy
centroid $m_1/m_0$ where the moments $m_1$ and $m_0$ were taken in a
certain energy interval around the resonance peak.  These windows are
$11<E^*<40$ MeV for GMR and GQR in $^{16}$O, $15<E^*<30$ MeV for the
GDR in $^{16}$O, $10 < E < 30$ MeV for GMR in $^{40, 48}$Ca, and $10 <
E < 25$ MeV for GQR in $^{40, 48}$Ca, The centroids $E_0$ for the GDR
in 40, 48Ca and for the GDR, GMR, and GQR in 208Pb were calculated in
the window $E_0 \pm 2\delta $ where $\delta$ is the spectral
dispersion (although with constraint $\delta \ge 2$MeV).
\begin{table}
\begin{center}
\begin{tabular}{|l|ccc|ccc|ccc|}
\hline
      &\multicolumn{3}{|c}{GDR} &\multicolumn{3}{|c}{GMR}
      &\multicolumn{3}{|c|}{GQR}\\
    &  exp. & RPA & TBA  &  exp. & RPA & TBA  &  exp. & RPA & TBA \\
\hline
$^{16}$O  &  24.3& 20.8& 19.7&   21.1&  23.1&  22.4&   19.8&   20.1&  19.9\\
$^{40}$Ca &  20.4& 19.0& 17.3&   18.7&  21.1&  20.4&   17.3&   16.6&  16.3\\
$^{48}$Ca &  20.2& 19.3& 18.5&   19.0&  20.5&  20.0&   16.6&   16.8&  16.3\\
$^{208}$Pb&  13.5& 14.3& 13.6&   13.8&  14.0&  13.8&   11.5&   10.9&  10.4\\
\hline
\end{tabular}
\end{center}
\caption{\label{tab:peaks}
Peak energies for GDR, GMR, abd GQR in four doubly-magic nuclei
computed with then parametrization SV-bas.
Compared are RPA an TBA results with the experimental value.
}
\end{table}
Table \ref{tab:peaks} shows the results for the four nuclei and three
resonance modes under consideration. We do that for one
parametrization, SV-bas, only in order to concentrate on the trends
with system size. The difference between RPA and TBA for the GDR is
about the same for all four nuclei while the GMR shows a significant
increase towards smaller nuclei and the GQR has the opposite trend to
yield smaller difference for smaller systems.  In any cases, the trends
are not nearly as strong as they were in earlier calculations, see
e.g. \cite{Lyutorovich_2012}. The subtraction scheme (\ref{Wsubtract})
tends to reduce the shift of resonance energies while maintaining full
collisional broadening from phonon coupling. In the terms of many-body
theory this means that the subtraction scheme reduces the effect on
the real part of the resonance energy while maintaining the full effect
on the imaginary part \cite{Toe88a}.  What the comparison with
experimental results is concerned, we see acceptable agreement for
$^{208}$Pb. That is the nucleus where SV-bas was tuned to the
resonances. Significant differences develop for lighter nuclei. This
is a known problem for the GDR \cite{Erler_2010}. A thorough study of
the $A$-dependence for the isoscalar modes has still to come.

\begin{figure}
\centerline{\includegraphics[width=\linewidth]{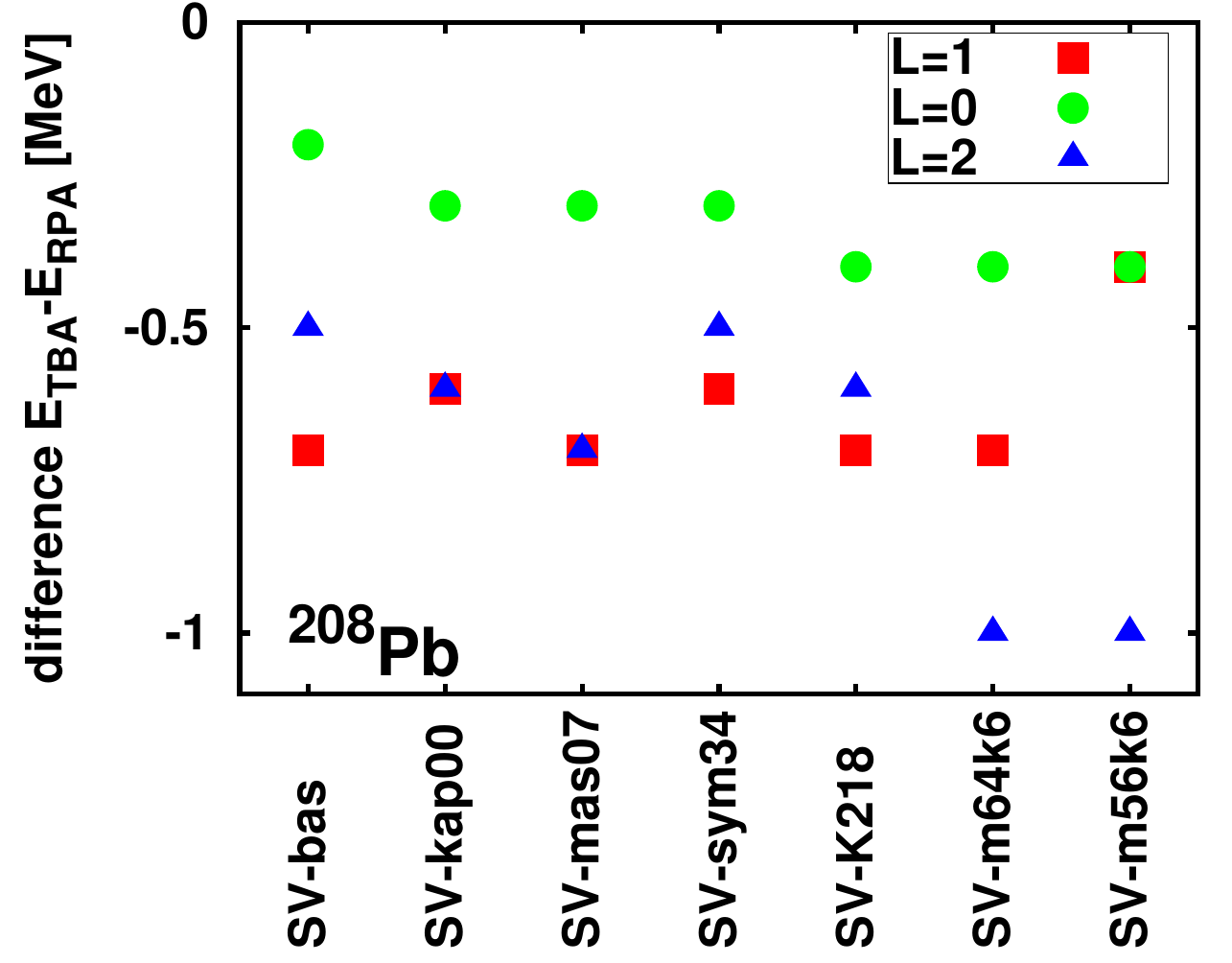}
}
\caption{\label{fig:diff-208Pb} Difference between RPA and TBA
  for the average energies of the three giant resonance modes in
  $^{208}$Pb for a variety of Skyrme parametrizations as indicated.
}
\end{figure}
Figure \ref{fig:diff-208Pb} shows the difference between RPA and TBA
resonance energies for one nucleus $^{208}$Pb, but now for all
parametrizations under consideration. The GMR seems rather robust as
it shows a small correction with little dependence on the
parametrization. Sizable shifts are seen for GDR as well as GQR and
these can vary significantly with parametrization. For the GQR, the
trend is clear. The shift increases with decreasing effective mass
$m^*/m$ and shows also some influence from the isovector effective
mass characterized by $\kappa_\mathrm{TRK}$. The GDR, again, seems to
be rather robust, except for the one parametrization SV-m56k6 for which
many NMP had been changed.

\section{Summary}

We have reviewed recent progess of the TBA, a selfconsistent generalization of the Landau-Migdal theory
based on effective Skyrme interactions and incorporating the phonon degree of freedom.
Phonons are shown to be a relevant degree of freedom in theoretical approaches to the
nuclear multipole strength functions. In heavy nuclei, a major improvement with respect
to a mean-field approximation(RPA) is obtained.
The performance of the method deteriorates in light nuclei, however, where phonons
by themselves do not suffice to account for the experimentally observed fragmentation of the strength.
Other degrees of freedom appear to be important which limits the method to heavy nuclei.
As the method is based on the effective Skyrme interaction which is valid for all nuclei, 
extrapolations to heavy neutron-rich nuclei are possible.

\begin{acknowledgments}
This work has been supported by contract Re322-13/1from the DFG.
\end{acknowledgments}

\appendix
\newcommand{\half}{{\textstyle\frac{1}{2}}}
\newcommand{\ffrac}[2]{{\textstyle\frac{#1}{#2}}}

\section{The Skyrme energy functional}
\label{app:SHF}

\subsection{Basic constituents: Densities and currents}
\label{sec:curr}

In SHF, a system is described in term of a set of single particle
(s.p.) wavefunctions $\varphi_\alpha(\vec{r})$ together with BCS
amplitudes $v_\alpha$ for occupation and
$u_\alpha=\sqrt{1-v_\alpha^2}$ for non-occupation.  
These are summarized in the one-body density matrix
\begin{equation}
  \varrho_q(\mathbf{r},\mathbf{r}')
  =
  \sum_{\alpha\in q} w_\alpha v_{\alpha}^2
     \varphi_{\alpha}^{\mbox{}}(\mathbf{r})\varphi_{\alpha}^\dagger(\mathbf{r}')
\label{eq:onebodyd}
\end{equation}
where $q$ labels the nucleon species with $q=p$ for protons and $q=n$
for neutrons. The $w_\alpha$ is a further factor which describes a
cutoff for pairing space (see below).  The SHF functional requires the
knowledge of only a few local densities and and currents, sorted here
according to time parity:
\begin{equation}
  \vspace*{-1em}
  \begin{array}{rcll}
  \multicolumn{3}{l}{\mbox{time even:}}
\\[2pt]
   \rho_q\hspace*{-0.5em}&=&\hspace*{-0.5em}\displaystyle
   \mbox{tr}_\sigma\{\varrho(\mathbf{r},\mathbf{r}')\}\big|_{r=r'}
  &\hspace*{-0.5em}\equiv\mbox{density}
\\
   \tau_q\hspace*{-0.5em}&=&\hspace*{-0.5em}\displaystyle
   \mbox{tr}_\sigma\{\nabla_r\nabla_{r'}\varrho(\mathbf{r},\mathbf{r}')\}\big|_{r=r'}
  &\hspace*{-0.5em}\equiv\mbox{kinetic density}
  \\
   \vec{J}_q \hspace*{-0.5em}&=&\hspace*{-0.5em}\displaystyle
   -\mathrm{i}\,\mbox{tr}_\sigma\{\nabla_r\!\times\!\hat{\vec{\sigma}}
         \varrho(\mathbf{r},\mathbf{r}')\}\big|_{r=r'}
  &\hspace*{-0.5em}\equiv\mbox{spin-orbit density}
\\[14pt]
  \multicolumn{3}{l}{\mbox{time odd:}}
\\[2pt]
   \vec{\sigma}_q\hspace*{-0.5em}&=&\displaystyle\hspace*{-0.5em}
   \mbox{tr}_\sigma\{\hat{\vec{\sigma}}\varrho(\mathbf{r},\mathbf{r}')\}\big|_{r=r'}
  &\hspace*{-0.5em}\equiv\mbox{spin density}
  \\
   \vec{j}_q\hspace*{-0.5em}&=&\hspace*{-0.5em}\displaystyle
   \Im\left\{\mbox{tr}_\sigma\{\nabla_r\varrho(\mathbf{r},\mathbf{r}')\}\big|_{r=r'}\right\}
  &\hspace*{-0.5em}\equiv\mbox{current}
\\
   \vec{\tau}_q\hspace*{-0.5em}&=&\hspace*{-0.5em}\displaystyle
   -\mbox{tr}_\sigma\{\hat{\vec{\sigma}}\nabla_r\nabla_{r'}
                       \varrho(\mathbf{r},\mathbf{r}')\}\big|_{r=r'}
  &\hspace*{-0.5em}\equiv\mbox{kinetic spin-dens.}
\\[14pt]
  \multicolumn{3}{l}{\mbox{time mixed:}}
\\[2pt]
   \xi_q 
   \hspace*{-0.5em}&=&\hspace*{-0.5em}
     \sum_{\alpha\in q}w_\alpha  u_{\alpha}v_{\alpha}|\varphi_{\alpha}|^2
  &\hspace*{-0.5em}\equiv\mbox{pairing density}
  \end{array}
\label{eq:rtj}
\vspace*{1em}
\end{equation}
It is advantageous to handle the densities in terms of isospin
$T\in\{0,1\}$ instead of protons $p$ and neutrons $n$. Thus we 
consider the recoupled forms which read for the local density
\begin{equation}
  \rho_{0}
  \equiv
  \rho
  =
  \rho_p+\rho_n
  \quad,\quad
  \rho_{1}
  =
  \rho_p-\rho_n
  \quad,
\end{equation}
and similarly for the other densities and currents.  The isoscalar
density $\rho_{0}\equiv\rho$ is equivalent to the total density and
the difference $\rho_{1}$ corresponds to the isovector density.
All densities and currents in the collection (\ref{eq:rtj}) 
are real and have definite time parity, except for the pairing
density $\xi$ which is complex and has mixed time parity.

\subsection{The energy functional}

The total energy in SHF is composed from kinetic energy, Skyrme
interaction energy, Coulomb energy, pairing energy, and correlation
energy from low-energy collective modes, usually a center-of-mass and
a rotational correlation. It reads
\begin{subequations}
\label{eq:basfunct}
\begin{eqnarray}
  E_\mathrm{total}
  &=&  
  \int \! d^3 r \; \left\{\mathcal{E}_\mathrm{kin}
  + \mathcal{E}_\mathrm{Sk}\right\}
\nonumber\\
  &&
  + E_\mathrm{Coul}
  + E_\mathrm{pair}
  - E_\mathrm{corr}
  \quad,
\label{eq:Etot}
\\
  \mathcal{E}_\mathrm{kin} 
  &=&
  \frac{\hbar^2}{2m_p} \tau_p+  \frac{\hbar^2}{2m_n} \tau_n
  \quad.
\label{eq:ekin}
\\
  E_\mathrm{Coul}
  &=&
  \frac{e^2}{2}\int d^3r\,d^3r'
   \frac{\rho_p(\vec{r})\rho_p(\vec{r}')}{|\vec{r}-\vec{r}'|}
\\
   &&
   +
   \frac{3e^2}{4}\left(\frac{3}{\pi}\right)^{1/3}
   \int d^3r[\rho_p(\vec{r})]^{4/3}
   \quad.
\end{eqnarray}
The pairing energy $E_\mathrm{pair}$ is irrelevant for the
present applications to doubly-magic nuclei. The
correlation energy $E_\mathrm{corr}$ amounts to a center-of-mass
correction to ground-state observables and is ignored for
the RPA excitations. The key piece is the Skyrme energy-density
functional $\mathcal{E}_\mathrm{Sk}$ which can be sorted into
time-even and time-odd couplings as
\begin{equation}
\label{eq:ESkeven}
\begin{array}{rclcl}
  \mathcal{E}_\mathrm{Sk,even}
  &=&
  {C_0^\rho\,\rho_0^2}
  &+&
  {C_1^\rho\,\rho_1^2}
  \\
  &&
  +{C_0^{\rho,\alpha}\,\rho_0^{2+\alpha}}
  &+&
  C_1^{\rho,\alpha}\,\rho_1^2\rho_0^\alpha
  \\
  &&
  +{C_0^{\Delta\rho}\,\rho_0\Delta\rho_0}
  &+&
  C_1^{\Delta\rho}\,\rho_1\Delta\rho_1
  \\
  &&
  +{C_0^{\nabla J}\,\rho_0\nabla\!\cdot\!\vec{J}_0}
  &+&
  C_1^{\nabla J}\,\rho_1\nabla\!\cdot\!\vec{J}_1
  \\[3pt]
  &&
  +C_0^{\tau}\,\rho_0\tau_0
  &+&
  C_1^{\tau}\,\rho_1\tau_1
  \\[3pt]
  &&
  +C_0^{J}\,\vec{J}_0^2
  &+&
  C_1^{J}\,\vec{J}_1^2
\end{array}
\end{equation}
\begin{equation}
\label{eq:ESkodd}
\begin{array}{rclcl}
  \mathcal{E}_\mathrm{Sk,odd}
  &=&
  \textcolor{white}{+}C_0^{\sigma}\,\vec{\sigma}_0^2
  &+&
  C_1^{\sigma}\,\vec{\sigma}_1^2
  \\
  &&
  +C_0^{\sigma,\alpha}\,\vec{\sigma}_0^2\rho_0^\alpha
  &+&
  C_1^{\sigma,\alpha}\,\vec{\sigma}_1^2\rho_0^\alpha
  \\
  &&
  +C_0^{\Delta\sigma}\,\vec{\sigma}_0\Delta\vec{\sigma}_0
  &+&
  C_1^{\Delta\sigma}\,\vec{\sigma}_1\Delta\vec{\sigma}_1
  \\
  &&
  +{C_0^{\nabla J}}
             \,\mathbb{\vec{\sigma}}_0\!\cdot\!\nabla\!\times\!\vec{j}_0
  &+&
  C_1^{\nabla J}\,\vec{\sigma}_1\!\cdot\!\nabla\!\times\!\vec{j}_1
  \\[3pt]
  &&
  -C_0^{\tau}\,\vec{j}_0^2
  &-&
  C_1^{\tau}\,\vec{j}_1^2
  \\[3pt]
  &&
  -\half {C_0^{J}}\vec{\sigma}_{0} \!\cdot\! \vec{\tau}_{0} 
  &-&
  \half {C_1^{J}}\vec{\sigma}_{1} \!\cdot\! \vec{\tau}_{1} 
\end{array}
\end{equation}
\end{subequations}
Note that at some places the same coupling constants appear in
time-even and time-odd part. That serves to guarantee Galilean
invariance of the functional, proven for the case of equal nucleon
masses in \cite{Engel_1975}. Spin couplings
$\propto C_T^{\sigma}, C_T^{\sigma\alpha},C_T^{\Delta\sigma}$ play no
role for the natural-parity modes considered here and are only listed
for completeness.

\subsection{The Skyrme ``force''}

Originally SHF was proposed as an effective interaction \cite{Sky59a},
later on coined as ``Skyrme force''. In this approach, the SHF energy
functional is given as the expectation value of the Skyrme interaction
$\hat{V}_{Skyrme}$ for a Slater state $|\Phi\rangle$ (more generally a
BCS state), i.e.
\begin{equation*}
  \int\!d^3r\;\mathcal{E}_\mathrm{Sk}
  =
 \langle\Phi|\hat{V}_{Skyrme}|\Phi\rangle
\end{equation*}
with 
\begin{eqnarray}
  \hat{V}_{Skyrme}&=& \hat{V}^m + \hat{V}^{LS} + \hat{V}^t
\label{eq:SHFforce}\\[0.2cm]
\hat{V}^m &=&
  t_0(1\!+\!x_0 \hat{ P}_\sigma)\delta(\mathbf{r}_{12})
\nonumber\\
  &&
  +
   \frac{t_3}{6}(1\!+\!x_3\hat P_\sigma)
  \rho^\alpha\left(\mathbf{r}_1\right)
  \delta(\mathbf{r}_{12})
\nonumber\\
  &&
  \!+ \frac{t_1}{2}(1\!+\!x_1\hat{P}_\sigma)
  \left(
   \delta(\mathbf{r}_{12})\hat{\boldsymbol k}^2
   +
   {\hat{\boldsymbol{k}}}'^{2}\delta(\mathbf{r}_{12})
  \right)
\nonumber\\
  &&
  + t_2(1\!+\!x_2\hat P_\sigma)\hat{\boldsymbol k}'
  \delta(\mathbf{r}_{12})\hat{\boldsymbol k}
\nonumber\\[0.2cm]
\hat{V}^{LS}&=&
  \mathrm{i}(W_0(1+x_WP_{\tau})(\hat{\boldsymbol{\sigma}} _1 + \hat{\boldsymbol{\sigma}}_2)
  \cdot \hat{\boldsymbol k}' \times \delta(\mathbf{r}_{12})
  \hat{\boldsymbol k}
\nonumber\\[0.2cm]
  \mathbf{r}_{12}
  &=&
  \textbf r_1-\textbf r_2
  \quad,\quad
  \hat P_\sigma
  =
  \frac{1}{2}(1+\hat{\boldsymbol \sigma}_1\hat{\boldsymbol{\sigma}}_2)
  \quad,\quad
\nonumber\\
  \hat{\boldsymbol k} 
  &=&
  -\frac{i}{2 }\left(\stackrel{\rightarrow}{\boldsymbol\nabla}_1
           -\stackrel{\rightarrow}{\boldsymbol\nabla}_2\right)
  \quad,\quad
\nonumber\\
  \hat{\boldsymbol k}'
  &=&
  \frac{i}{2 }\left(\stackrel{\leftarrow}{\boldsymbol\nabla}_1
      -\stackrel{\leftarrow}{\boldsymbol\nabla}_2\right)
  \quad.
\nonumber
\end{eqnarray}
where $\hat{\boldsymbol k}$ acts to the right and $\hat{\boldsymbol
  k}'$ to the left.  Note that the rather involved, but usually
ignored, tensor force has not been listed here.  This ``Skyrme force''
was motivated in that form from a low momentum expansion of the
density-matrix \cite{Neg72a,Bog10aR} which yields the above terms
where each parameter $t_i$ carries, in principle, some density
dependence. For simplicity, one has decided to ignore density
dependence, except for the zeroth order term which is augmented by a
density dependence $\rho^\alpha$.  We put the notion ``force'' in
quotation mark because this object depends on the density which is
produced by the wave function on which this force acts. This is not a
standard two-body operator, but an effective force designed for
building an expectation value with a mean-field state
\cite{Erler_2010}.
 
Each one of the two viewpoints, the energy functional
(\ref{eq:basfunct}) and the Skyrme ``force'' (\ref{eq:SHFforce}), has
a natural set of parameters associated to the terms of the functional
(the $C_T^\mathrm{(typ)}$), or force (the $t_i$) respectively.  There
is a one-to-one correspondence between the two conventions: 
\begin{equation}
  \hspace*{-4.5em}
  \begin{array}{rl}
  C_0^\rho &  = \ffrac{3}{8}  t_0 \;,
\\[2pt]         
  C_1^\rho & = -\ffrac{1}{4}  t_0 (\half + x_0)   \;, 
\\[2pt]
  C_0^\tau & = \ffrac{3}{16} t_1 + \ffrac{5}{16} t_2 + \ffrac{1}{4} t_2 x_2
         \;, 
\\[2pt]
  C_1^\tau & = -\ffrac{1}{8} \Big[ t_1 (\half+x_1)-t_2 (\half+x_2)\Big] 
       \;, 
\\[2pt]
  C_0^{\Delta \rho} &  =   -\ffrac{9}{64} t_1  +\ffrac{5}{64} t_2
            +\ffrac{1}{16}  t_2 x_2
       \;,
\\[2pt]
  C_1^{\Delta \rho} &
         = \ffrac{1}{32} \Big[ 3t_1 (\half+x_1)+t_2 (\half+x_2)\Big] 
         \;, 
\\[2pt]
  C_0^{\rho\alpha} &  = \ffrac{1}{16} t_3
    \;, 
\\[2pt]
  C_1^{\rho\alpha} &
       = -\ffrac{1}{24} t_3 (\half+x_3)
   \;,
\\[2pt]
  C_0^{\nabla J} &   = -\ffrac{3}{4}t_4
  \;
\\[2pt]
  C_1^{\nabla J} &  = -\ffrac{1}{4} t_4
         \; , 
\\[2pt]
  C_0^s & = - \ffrac{1}{4} t_0 \big( \half - x_0 \big)
         \;, 
\\[2pt]
  C_1^s &=  -\ffrac{1}{8} t_0  \;, 
\\[2pt]
  C_0^{sT} &= -\ffrac{1}{8} \Big[   t_1 \big( \half - x_1 \big)
                            - t_2 \big( \half + x_2 \big)
                      \Big]
         \;, 
\\[2pt]
  C_1^{sT} & = - \ffrac{1}{16} ( t_1 - t_2 )
          \;, 
\\[2pt]
  C_0^{\Delta s} &
    =  \ffrac{1}{32} \Big[ 3 t_1 \big( \half - x_1 \big)
                               + t_2 \big( \half + x_2 \big) 
                         \Big]           
     \;,
\\[2pt]
  C_1^{\Delta s} &= \ffrac{1}{64} ( 3 t_1 + t_2 ) 
         \; , 
\\[2pt]
  C_0^{s\alpha} &  = - \ffrac{1}{24} t_3 (\half - x_3)
    \;,
\\[2pt]
  C_1^{s\alpha} &  = -\ffrac{1}{48} t_3
          \;, 
  \end{array}
\label{eq:bdef2}
\end{equation}
There is one exception concerning the spin-orbit term. The energy
functional (\ref{eq:basfunct}) allows for independent choice of
$C_0^{sT}$ and $C_1^{sT}$.  That is freedom which was recommended in
\cite{RF95} and regularly used later on.  But the force
(\ref{eq:SHFforce}) ties these two terms together as seen in
eq. (\ref{eq:bdef2}).

The derivation of the reduced elements of RPA residual two-body
interaction in appendix \ref{append1} refers to the force aspect of
SHF and thus formulates the matrix elements preferably in terms of the
force parameters $t_i$.

\subsection{Nuclear matter parameters}

Infinite nuclear matter is taken without Coulomb force, pairing, and
correlation correction. It remains the energy per particle as
\begin{equation}
  \frac{E}{A}(\rho_0,\rho_1,\tau_0,\tau_1)
  = 
  \frac{\mathcal{E}_\mathrm{kin}+\mathcal{E}_\mathrm{Sk}}{\rho_0}
  \quad.
\end{equation}
where we consider for a while $\rho$ and $\tau$ as independent
variables. Of course, a given system is characterized just
by the densities $\rho_T$ while the kinetic density depends on these
given densities as $\tau_T=\tau_T(\rho_0,\rho_1)$. Thus we have to
distinguish between partial derivatives $\partial/\partial_\tau$ which
take $\tau_T$ as independent and total derivatives $d/d\rho$ which
deal only with $\rho_T$ dependence. The relation is
\begin{equation}
  \frac{d}{d\rho_T}
  =
  \frac{\partial}{\partial\rho_T}
  +
  \sum_{T'}\frac{\partial\tau_{T'}}{\partial\rho_T}
  \frac{\partial}{\partial\tau_{T'}}
  \quad.
\end{equation}
The standard NMP are defined at the equilibrium point
($\rho_0=\rho_\mathrm{eq}$, $\rho_1=0$) of symmetric nuclear matter.
\begin{table}[t]
\begin{center}
\begin{tabular}{lrcl}
\hline
\multicolumn{4}{c}{\rule{0pt}{12pt}isoscalar ground state properties}
\\[4pt]
\hline
 equilibrium density:\hspace*{-0.5em}
 &\rule{0pt}{16pt}
 $\rho_\mathrm{eq}$
 &$\leftrightarrow$&
 $\displaystyle\frac{d}{d\rho_0}\frac{E}{A}\Big|_\mathrm{eq}=0$
\\[12pt]
 equilibrium energy:\hspace*{-0.5em}
 &
 $\displaystyle\frac{E}{A}\Big|_\mathrm{eq}$
\\[12pt]
\hline
\multicolumn{4}{c}{\rule{0pt}{12pt}isoscalar response properties}
\\[4pt]
\hline
 incompressibility:\hspace*{-0.5em}
 &
  $  K_\infty$
  &=& 
  $\displaystyle
  9\,\rho_0^2 \, \frac{d^2}{d\rho_0^2} \,
       \frac{{E}}{A}\Big|_\mathrm{eq}
  $
\\[12pt]
  effective mass:\hspace*{-0.5em}
  &
  $\displaystyle
  \frac{\hbar^2}{2m*}
  $
  &=&
  $\displaystyle
   \frac{\hbar^2}{2m}
    + 
    \frac{\partial}{\partial\tau_0} \frac{{E}}{A}\bigg|_\mathrm{eq}
  $
\\[14pt]
\hline
\multicolumn{4}{c}{\rule{0pt}{12pt}isovector response properties}
\\[4pt]
\hline
  symmetry energy:\hspace*{-0.5em}
  &
  $J$
  &=&
  $\displaystyle\rule{0pt}{19pt}
 \frac{1}{2} \rho_0^2\frac{d^2}{d\rho_1^2}
  \frac{{E}}{A} \bigg|_\mathrm{eq}
  $
\\[12pt]
  slope of $J$:\hspace*{-0.5em}
  &
  $L$
  &=&
  $\displaystyle
  {3}\rho_0 \frac{d}{d\rho_0}J
  $
\\[12pt]
  TRK sum-rule enh.:\hspace*{-0.5em}
  &
  $\kappa_{\rm TRK}$
  &=& 
  $\displaystyle
  \frac{2m}{\hbar^2}
  \frac{\partial}{\partial\tau_1} 
  \frac{{E}}{A}\bigg|_\mathrm{eq}
  $
\\[12pt]
\hline
\end{tabular}
\end{center}
\caption{\label{tab:nucmatdef}
Definition of the nuclear matter properties (NMP).
All derivatives are to be  taken at the equilibrium point
of  symmetric nuclear matter. For the definition of $L$, $J$ is
considered temporarily as $J(\rho_0)$.
}
\end{table}
They are summarized in table \ref{tab:nucmatdef}.
The enhancement factor for the Thomas-Reiche-Kuhn (TRK) sum
rule \cite{Rin80aB} is a widely used way to characterize the isovector
effective mass which is obvious from the given expression involving
derivative with respect to $\tau_1$.  The slope of symmetry energy $L$
characterizes the density dependence of the symmetry energy which
allows to estimate the symmetry energy at half density, i.e. at
surface of finite nuclei.

The NMP 
can be grouped into four classes:
first, the (isoscalar) ground state properties $\rho_\mathrm{eq}$ and
${E}/{A}\Big|_\mathrm{eq}$, second, isoscalar response properties $K$
and $m/m$, and third, isovector response properties $J$, $L$,
$\kappa_\mathrm{TRK}$. The response properties determine zero sound in
matter \cite{Tho61aB} and subsequently they are closely related to
giant resonance modes in finite nuclei as we will see later.  There is
a further category, the surface energies which go already beyond
homogeneous matter and whose definition is rather
involved \cite{Rei06a}. They are not considered here.

Homogeneous matter yields $\Delta\rho=0$ and $\vec{J}=0$ which, in
turn, renders four terms in the functional\ref{eq:ESkeven} inactive.
Thus we have exactly seven interaction parameters ($C_0^{\rho}$,
$C_0^{\rho,\alpha}$, $C_0^{\tau}$, $\alpha$) to determine seven NMP.
The relation is revertible establishing a one-to-one correspondence between the
both sets. This allows to consider the NMP equivalently as model
parameters which is, in fact, a more intuitive way to communicate the model
parameters. And this is the aspect which is used in the systematic
variation of the SHF functional (see section \ref{sec:Skyrme}).


\subsection{Reduced matrix elements of the residual interaction}

\label{append1}

In this Appendix we draw the exact formulas for the reduced matrix elements
of the residual interaction $V$ deduced from the Skyrme energy functional
of the standard form (see, e.g., \cite{CBHMS98,Bender_2003}).

It is convenient to present this interaction as a sum of the following
terms
\be
{V}^{\vphu}_{12,34} = {V}^{\,(0)}_{12,34} +
{V}^{\,(2)D}_{12,34} + {V}^{\,(2)X}_{12,34} +
{V}^{\,(2)SO}_{12,34} + {V}^{\,(C)}_{12,34}\,,
\label{rmeri1}
\ee
where ${V}^{\,(0)}_{12,34}$ is the momentum-independent part of $V$
(including density-dependent terms),
${V}^{\,(2)D}_{12,34}$, ${V}^{\,(2)X}_{12,34}$ and ${V}^{\,(2)SO}_{12,34}$
are the direct, exchange and spin-orbit terms of the momentum-dependent part
of the interaction (all these terms are of the second order in the momenta,
see Ref.~\cite{NPA928} for the explicit definition of ${V}^{\,(2)D}_{12,34}$
and ${V}^{\,(2)X}_{12,34}$),
${V}^{\,(C)}_{12,34}$ is the Coulomb interaction.
%
It is supposed that the matrix elements in Eq. (\ref{rmeri1})
are calculated in the representation of the single-particle
wave functions $\vphi^{\vphu}_1(\bfr,\sigma,\tau)$
of some spherically-symmetric basis. In this case the matrix indices
can be represented as $1=\{(1),m_1\}$, where
$(1)=\{\tau_1,n_1,l_1,j_1\}$,
$m_1$ is a projection of the total angular momentum.
Let us define reduced matrix elements as follows
\bea
{V}^{\,J}_{(12,34)} &=&
\sum_{m_{\mbts{1}} m_{\mbts{2}} m_{\mbts{3}} m_{\mbts{4}} M}
{V}^{\vphu}_{12,34}
\nonumber\\
&\times&
(-1)^{\,j_{\mbts{2}} - m_{\mbts{2}}}\,
\left(
\begin{array}{ccc}
j_1 & j_2 & J \\ m_1 & -m_2 & M \\
\end{array}
\right)
\nonumber\\
&\times&
\,(-1)^{\,j_{\mbts{4}} - m_{\mbts{4}}}\,
\left(
\begin{array}{ccc}
j_3 & j_4 & J \\ m_3 & -m_4 & M \\
\end{array}
\right).
\label{defrme}
\eea
%
For the excitations in the neutral particle-hole channel
with the total angular momentum $J$ one obtains
\bea
{V}^{\,J}_{(12,34)} &=& {V}^{\,J\,(0)}_{(12,34)}
+ {V}^{\,J\,(2)D}_{(12,34)} + {V}^{\,J\,(2)X}_{(12,34)}
\nonumber\\
&+&
{V}^{\,J\,(2)SO}_{(12,34)} + {V}^{\,J(C)}_{(12,34)}\,.
\label{rmeri2}
\eea
Below the explicit formulas for the terms in this equality are
given.

\subsection{Momentum-independent part of the interaction}
\label{append1a}

\bea
&&{V}^{\,J\,(0)}_{(12,34)} =
\frac{\delta_{\tau_1,\tau_2}\,\delta_{\tau_4,\tau_3}}{2J+1}
\sum_{LS}\,{I}^{S(0)}_{(12,34)}
\nonumber\\
&&\times\,
\langle\,j_2 l_2\,||\,T_{JLS}\,||\,j_1 l_1 \rangle\,
\langle\,j_4 l_4\,||\,T_{JLS}\,||\,j_3 l_3 \rangle\,,
\label{rmevz}
\eea
where $S=0,1$,
\bea
\hspace{-2em}
&&{I}^{S(0)}_{(12,34)} = \int_0^{\infty} dr\,r^2\,
\bigl[\,
\delta^{\vphu}_{S,\,0}\,F^{\,0}_{\,\tau_1,\,\tau_3}(r) +
\delta^{\vphu}_{S,\,1}\,G^{\,0}_{\,\tau_1,\,\tau_3}(r)\,
\bigr]
\nonumber\\
\hspace{-2em}
&&\times\,
R_{(1)}(r)\,R_{(2)}(r)\,R_{(3)}(r)\,R_{(4)}(r)\,,
\label{defis}
\eea
$R_{(1)}(r)$ is the radial part of the single-particle wave
function $\vphi^{\vphu}_1(\bfr,\sigma,\tau)$.
The functions
$F^{\,0}_{\,\tau,\,\tau'}(r)$ and $G^{\,0}_{\,\tau,\,\tau'}(r)$
are defined by the following equations
\bea
&&F^{\,0}_{n,n} = \frac{1}{2}\,(1 - x_0)\,t_0
+\frac{t_3}{48}\,\rho^{\alpha}\,\biggl\{
3(\alpha + 1)(\alpha + 2)
\nonumber\\
&&-\; (1 +2\,x_3)\,
\bigl[\,\alpha(\alpha -1)(\bar{\rho}/\rho)^2
+ 4\alpha\,\bar{\rho}/\rho + 2\,\bigr]\biggr\}\,,
\label{deff0nn}
\eea
\bea
&&F^{\,0}_{n,p} = F^{\,0}_{p,n} = (\,1 + x_0/2\,)\,t_0
+ \frac{t_3}{48}\,\rho^{\alpha}\,\biggl\{
3(\alpha + 1)(\alpha + 2)
\nonumber\\
&&-\; (1 +2\,x_3)\,
\bigl[\,\alpha(\alpha -1)(\bar{\rho}/\rho)^2 - 2\,
\bigr]\biggr\}\,,
\label{deff0np}
\eea
\bea
%
G^{\,0}_{n,n} &=& - \frac{1}{2}\,(1 - x_0)\,t_0
- \frac{t_3}{12}\,(1 - x_3)\,\rho^{\alpha},
\label{defg0nn}\\
\vphantom{A^{\ds B^{\ds C^{\ds D}}}}
G^{\,0}_{n,p} &=& G^{\,0}_{p,n}\,=\;\frac{x_0\,t_0}{2}
+ \frac{x_3\,t_3}{12}\,\rho^{\alpha},
\label{defg0np}
\eea
where
$t_0$, $x_0$, $t_3$, $x_3$, and $\alpha$ are the parameters
of the Skyrme energy functional (see, e.g., \cite{CBHMS98}),
$\rho = \rho^{\vphu}_n + \rho^{\vphu}_p\,$,
$\,\bar{\rho} = \rho^{\vphu}_n - \rho^{\vphu}_p\,$,
$\rho^{\vphu}_n = \rho^{\vphu}_n(r)$ and
$\rho^{\vphu}_p = \rho^{\vphu}_p(r)$
are the neutron and proton local densities.
Formulas for $F^{\,0}_{p,p}$ and $G^{\,0}_{p,p}$ are
obtained from Eqs.~(\ref{deff0nn}) and (\ref{defg0nn})
by replacing $\bar{\rho} \rightarrow -\bar{\rho}$.
$\langle\,j_1 l_1\,||\,T_{JLS}\,||\,j_2 l_2 \rangle$
is the reduced matrix element of the spherical tensor operator
$T_{JLSM}=(Y_L\otimes \sigma_S)_{JM}$
which is defined by the formula (see also Eqs. (A1)--(A4) of
Ref.~\cite{Litvinova_2007})
\bea
&&\langle\,j_1 l_1\,||\,T_{JLS}\,||\,j_2 l_2 \rangle
\nonumber\\
&&=\,
(-1)^{l_{\mbts{1}}} \sqrt{\frac{(2J+1)(2L+1)(2S+1)}{2\pi}}
\nonumber\\
&&\times\,
\sqrt{(2j_1+1)(2l_1+1)(2j_2+1)(2l_2+1)}\,
\nonumber\\
&&\times
\left(
\begin{array}{ccc}
l_1 & l_2 & L \\ 0 & 0 & 0 \\
\end{array}
\right)
\left\{
\begin{array}{ccc}
\frac{1}{2} & l_2 & j_2 \\ \frac{1}{2} & l_1 & j_1 \\
S & L & J \\
\end{array}
\right\}.
\label{deftjls}
\eea

\subsection{Direct and exchange terms of the
momentum-dependent part of the interaction}
\label{append1b}

\bea
&&{V}^{\,J\,(2)D}_{(12,34)} =
\frac{\delta_{\tau_1,\tau_2}\,\delta_{\tau_4,\tau_3}}{2J+1}
\sum_{S}\,C^{(S)D}_{\tau_1,\,\tau_3}\,{U}^{JS}_{(12,34)}\,,
\label{rmevd}\\
\vphantom{A^{\ds B^{\ds C^{\ds D}}}}
&&{V}^{\,J\,(2)X}_{(12,34)} =\,
\delta_{\tau_1,\tau_2}\,\delta_{\tau_4,\tau_3}
\sum_{SJ'}\,C^{(S)X}_{\tau_1,\,\tau_3}\,
{U}^{J'S}_{(42,31)}
\nonumber\\
&&\times\,
(-1)^{J+J^{\,\prime}+j_{\mbts{1}}-j_{\mbts{4}}}\,
\left\{
\begin{array}{ccc}
j_2 & j_4 & J^{\,\prime} \\ j_3 & j_1 & J \\
\end{array}
\right\},
\label{rmevx}
\eea
where $S=0,1$,
\bea
\hspace{-1em}
&&{U}^{JS}_{(12,34)} =
- \sum_{L}\,I^{L(2)}_{(12,34)}
\nonumber\\
\hspace{-1em}
&&\times\,
\langle\,j_2 l_2\,||\,T_{JLS}\,||\,j_1 l_1 \rangle\,
\langle\,j_4 l_4\,||\,T_{JLS}\,||\,j_3 l_3 \rangle\,,
\label{rmeusj}
\eea
\bea
&&I^{L(2)}_{(12,34)} = \int_0^{\infty} dr\biggl[\,r^2\,
\bigl( R_{(1)}(r)\,R_{(2)}(r)\,\bigr)^{\prime}
\bigl( R_{(3)}(r)\,R_{(4)}(r)\,\bigr)^{\prime}
\nonumber\\
&&+\, L(L+1)\,
R_{(1)}(r)\,R_{(2)}(r)\,R_{(3)}(r)\,R_{(4)}(r)\,\biggr]\,,
\label{defil}
\eea
\bea
\hspace{-2em}
C^{(S)D}_{\tau,\,\tau'} &=&
- \bigl[\,b^{(-)}_{S0} +
(2\delta^{\vphu}_{\tau,\,\tau'} - 1)\,b^{(-)}_{S1}\bigr],
\label{defcds}\\
\hspace{-2em}
C^{(S)X}_{\tau,\,\tau'} &=& \frac{1}{2}\,
\bigl\{ b^{(+)}_{00} +
(2\delta^{\vphu}_{\tau,\,\tau'} - 1)\,b^{(+)}_{01}
\nonumber\\
\hspace{-2em}
&+& \bigl( 3 - 4S \bigr)
\bigl[\,b^{(+)}_{10} +
(2\delta^{\vphu}_{\tau,\,\tau'} - 1)\,b^{(+)}_{11}\bigr]
\bigr\}.
\label{defcxs}
\eea
Normally, the parameters $b^{(\pm)}_{ST}$ are expressed through
the ordinary Skyrme-force parameters $t_1$, $x_1$, $t_2$, $x_2$
by the equations
\bea
b^{(\pm)}_{00} &=& \frac{1}{16}\bigl[ \pm\,(5+4x_2)\,t_2 + 3\,t_1\bigr]\,,
\label{defc00}\\
b^{(\pm)}_{10} &=& \frac{1}{16}\bigl[ \pm\,(1+2x_2)\,t_2 - (1-2x_1)\,t_1\bigr]\,,
\label{defc10}\\
b^{(\pm)}_{01} &=& \frac{1}{16}\bigl[ \pm\,(1+2x_2)\,t_2 - (1+2x_1)\,t_1\bigr]\,,
\label{defc01}\\
b^{(\pm)}_{11} &=& \frac{1}{16}( \pm\;t_2 - t_1)\,.
\label{defc11}
\eea
If Eqs. (\ref{defc00})--(\ref{defc11}) are fulfilled, we have
\be
{V}^{\,(2)X}_{12,34} = -{V}^{\,(2)D}_{42,31} = -{V}^{\,(2)D}_{13,24}\,.
\label{xdsym}
\ee
However, in the general case there is another way of the choice
of these parameters in which they are expressed
through the coupling constants of the Skyrme energy functional
$C^{\tau}_T$, $C^{J}_T$, $C^{\Delta\rho}_T$, and $C^{\Delta s}_T$
(choice (ii) of Ref.~\cite{Bender_2003}). In this case we have
\bea
b^{(+)}_{0T} &=& \hphantom{-} C^{\tau}_T\,,
\label{bptc0}\\
b^{(+)}_{1T} &=& - C^{J}_T\,,
\label{bptc1}\\
b^{(-)}_{0T} &=& - 2\,C^{\Delta\rho}_T - \frac{1}{2}\,C^{\tau}_T\,,
\label{bmtc0}\\
b^{(-)}_{1T} &=& - 2\,C^{\Delta s}_T + \frac{1}{2}\,C^{J}_T\,.
\label{bmtc1}
\eea
The inverse formulas read
\bea
C^{\Delta\rho}_T &=&
- \frac{1}{2}\,b^{(-)}_{0T} - \frac{1}{4}\,b^{(+)}_{0T}\,,
\label{cdrho}\\
C^{\Delta s}_T &=&
- \frac{1}{2}\,b^{(-)}_{1T} - \frac{1}{4}\,b^{(+)}_{1T}\,,
\label{cdels}\\
C^{\tau}_T &=& \hphantom{-} b^{(+)}_{0T},
\label{ctaut}\\
C^{J}_T &=& - b^{(+)}_{1T}.
\label{cjbt}
\eea
In contrast to Eqs. (\ref{defc00})--(\ref{defc11}),
Eqs. (\ref{bptc0})--(\ref{bmtc1}) do not impose any constraints
on the parameters $b^{(\pm)}_{ST}$, because the numbers of the
independent parameters in the left and right sides of
Eqs. (\ref{bptc0})--(\ref{bmtc1}) are equal to each other.
In this case Eqs. (\ref{xdsym}) are generally not fulfilled.

The definitions (\ref{bptc0})--(\ref{bmtc1}) are convenient in the case
when it is necessary to eliminate the so-called $J^2$ terms
or/and the spin-spin terms from the residual interaction $V$.
In most parametrizations of the Skyrme energy functional the $J^2$ terms
are omitted by setting the constants $C^{J}_T$ to be equal to zero.
To maintain self-consistency on the RPA level these constants should
be equal to zero also in the residual interaction.
In this case the parameters $b^{(\pm)}_{ST}$ are determined
by Eqs. (\ref{bptc0})--(\ref{bmtc1}) in which $C^{J}_T = 0$ while
the coupling constants $C^{\Delta\rho}_T$, $C^{\Delta s}_T$, and
$C^{\tau}_T$ are determined by Eqs. (\ref{cdrho})--(\ref{ctaut}) and
(\ref{defc00})--(\ref{defc11}).
Note that in these definitions Eqs. (\ref{defc00})--(\ref{defc11}) play
intermediate role (they do not give the final values of the parameters
$b^{(\pm)}_{ST}$, so Eqs. (\ref{xdsym}) do not follow from them).

Sometimes in the RPA calculations of the excitations of the spherical
even-even nuclei the spin-spin terms of $V$ are also omitted since this
does not lead to the violation of the self-consistency.
In the above equations it means that (i) the sum in Eq.~(\ref{rmevz})
is restricted by the terms with $S=0$ and (ii) the constants
$C^{\Delta s}_T$ in Eq.~(\ref{bmtc1}) are set to be equal to zero
(the constants $C^{\tau}_T$, $C^{J}_T$, and $C^{\Delta\rho}_T$ are
determined as described above).

\subsection{Spin-orbit term of the
momentum-dependent part of the interaction}
\label{append1c}

\bea
&&{V}^{\,J\,(2)SO}_{(12,34)} =
\delta_{\tau_1,\tau_2}\,\delta_{\tau_4,\tau_3}
W_0\,(1 + x^{\vphuu}_W\,\delta_{\tau_1,\tau_3})
\nonumber\\
&&\times
\Bigl\{\,
u^{\;J}_{(12,34)} + u^{\;J}_{(34,12)}
\nonumber\\
&&+\;
(-1)^{\,j_{\mbts{1}} - j_{\mbts{2}} + j_{\mbts{3}} - j_{\mbts{4}}}\,
\bigl[\,u^{\;J}_{(21,43)} + u^{\;J}_{(43,21)}\,\bigr]
\Bigr\}\,,
\label{rmeso}
\eea
where
\bea
&&u^{\;J}_{(12,34)} =
\frac{3}{4\pi}\sum_{L = J, J \pm 1}
\sum_{\;\;l'_1 = l^{\vphd}_1 \pm 1} \sum_{\;\;l'_3 = l^{\vphd}_3 \pm 1}
\,a^{\,JLl'_1}_{(12)}\,b^{\,JLl'_3}_{(34)}
\nonumber\\
&&\times
\int_0^{\infty} dr\,r^2\,
D^{\,l'_1}_{(1)}(r)\,R^{\vphu}_{(2)}(r)
D^{\,l'_3}_{(3)}(r)\,R^{\vphu}_{(4)}(r)\,,
\label{defcll}
\eea
\bea
a^{\,JLl'_1}_{(12)} &=&
\sum_{l''_1 = j^{\vphd}_1 \pm \frac{1}{2}}\,(-1)^{\,l_{\mbts{1}} +\,l''_1}
(2l'_1+1)(2l''_1+1)
\nonumber\\
&\times&
\sqrt{(2j_1+1)(2l_1+1)(2j_2+1)(2l_2+1)}
\nonumber\\
&\times&
\left(
\begin{array}{ccc}
l^{\vphd}_1 & l'_1 & 1 \\ 0 & 0 & 0 \\
\end{array}
\right)
\left(
\begin{array}{ccc}
l^{\vphd}_2 & l'_1 & L \\ 0 & 0 & 0 \\
\end{array}
\right)
\nonumber\\
&\times&
\left\{
\begin{array}{ccc}
l^{\vphd}_1 & l'_1 & 1 \\ 1 & 1 & l''_1 \\
\end{array}
\right\}
\left\{
\begin{array}{ccc}
l^{\vphd}_2 & l'_1 & L \\ 1 & J & l''_1 \\
\end{array}
\right\}
\nonumber\\
&\times&
\left\{
\begin{array}{ccc}
j_1 & l_1 & \frac{1}{2} \\ 1 & \frac{1}{2} & l''_1 \\
\end{array}
\right\}
\left\{
\begin{array}{ccc}
j_2 & l_2 & \frac{1}{2} \\ l''_1 & j^{\vphd}_1 & J \\
\end{array}
\right\},
\label{defajll}
\eea
\bea
b^{\,JLl'_3}_{(34)} &=&
(-1)^{\,l_{\mbts{3}} + j_{\mbts{3}} - \frac{1}{2}}
\,(2L+1)\,(2l'_3+1)
\nonumber\\
&\times&
\sqrt{(2j_3+1)(2l_3+1)(2j_4+1)(2l_4+1)}
\nonumber\\
&\times&
\left(
\begin{array}{ccc}
l^{\vphd}_3 & l'_3 & 1 \\ 0 & 0 & 0 \\
\end{array}
\right)
\left(
\begin{array}{ccc}
l^{\vphd}_4 & l'_3 & L \\ 0 & 0 & 0 \\
\end{array}
\right)
\nonumber\\
&\times&
\left\{
\begin{array}{ccc}
j_4 & l_4 & \frac{1}{2} \\ l_3 & j_3 & J \\
\end{array}
\right\}
\left\{
\begin{array}{ccc}
l^{\vphd}_4 & l'_3 & L \\ 1 & J & l_3 \\
\end{array}
\right\},
\label{defbjll}
\eea
\bea
D^{\,l'_1}_{(1)}(r) &=& R^{\,\prime\vphu}_{(1)}(r)
- \sqrt{6 l_1(l_1+1)(2l_1+1)}
\nonumber\\
&\times&
\left\{
\begin{array}{ccc}
1 & l^{\vphd}_1 & l'_1 \\ l_1 & 1 & 1 \\
\end{array}
\right\}
\frac{1}{r}\,R^{\vphu}_{(1)}(r)\,.
\label{defdso}
\eea
$W_0$ and $x^{\vphuu}_W$ in Eq.~(\ref{rmeso})
are the parameters of the Skyrme energy functional
(see \cite{SLKR95}). Note that these parameters are related
with the constants $b^{\vphuu}_4$ and $b'_4$ of Ref.~\cite{RF95}
by the formulas
$W_0=2b_4$, $x^{\vphuu}_W=b'_4/b^{\vphuu}_4$.

\subsection{Coulomb term}
\label{append1d}

The Coulomb term has the non-zero matrix elements only for
the proton single-particle wave functions. It consists of two parts
\be
{V}^{\,J(C)}_{(12,34)} =
{V}^{\,J(C)D}_{(12,34)} + {V}^{\,J(C)X}_{(12,34)},
\label{rmec}
\ee
where ${V}^{\,J(C)D}_{(12,34)}$ and ${V}^{\,J(C)X}_{(12,34)}$
are the direct and exchange terms, respectively.

For the direct term we have
\bea
{V}^{\,J(C)D}_{(12,34)} &=& \frac{4\pi e^2}{(2J+1)^2}
\nonumber\\
&\times&
\langle\,j_2 l_2\,||\,T_{JJ0}\,||\,j_1 l_1 \rangle\,
\langle\,j_4 l_4\,||\,T_{JJ0}\,||\,j_3 l_3 \rangle
\nonumber\\
&\times&
\int_0^{\infty} dr\,r^2\int_0^{\infty} dr'\,r'^2\,
\frac{r_{<}^J}{r_{>}^{J+1}}
\nonumber\\
&\times&
R_{(1)}(r)\,R_{(2)}(r)\,R_{(3)}(r')\,R_{(4)}(r')\,,
\label{rmecd}
\eea
where $r_{<}^{\vphd}=\min\,(r,r')$, $r_{>}^{\vphd}=\max\,(r,r')$.

The exchange term is treated within the Slater approximation
in consistency with the usual form of the Skyrme energy functional.
In this approximation the expression for the exchange term is
\bea
{V}^{\,J(C)X}_{(12,34)} &=& -\frac{(9\pi)^{-1/3}e^2}{2J+1}
\nonumber\\
&\times&
\langle\,j_2 l_2\,||\,T_{JJ0}\,||\,j_1 l_1 \rangle\,
\langle\,j_4 l_4\,||\,T_{JJ0}\,||\,j_3 l_3 \rangle
\nonumber\\
&\times&
\int_0^{\infty} dr\,r^2\,\rho_p^{-2/3}(r)
\nonumber\\
&\times&
R_{(1)}(r)\,R_{(2)}(r)\,R_{(3)}(r)\,R_{(4)}(r)\,.
\label{rmecx}
\eea


\bibliographystyle{apsrev4-1}
\bibliography{GMQR,PGR}

\begin{thebibliography}{10}%
\makeatletter
\providecommand \@ifxundefined [1]{%
 \ifx #1\undefined \expandafter \@firstoftwo
 \else \expandafter \@secondoftwo
\fi
}%
\providecommand \@ifnum [1]{%
 \ifnum #1\expandafter \@firstoftwo
 \else \expandafter \@secondoftwo
\fi
}%
\providecommand \enquote [1]{``#1''}%
\providecommand \bibnamefont  [1]{#1}%
\providecommand \bibfnamefont [1]{#1}%
\providecommand \citenamefont [1]{#1}%
\providecommand\href[0]{\@sanitize\@href}%
\providecommand\@href[1]{\endgroup\@@startlink{#1}\endgroup\@@href}%
\providecommand\@@href[1]{#1\@@endlink}%
\providecommand \@sanitize [0]{\begingroup\catcode`\&12\catcode`\#12\relax}%
\@ifxundefined \pdfoutput {\@firstoftwo}{%
 \@ifnum{\z@=\pdfoutput}{\@firstoftwo}{\@secondoftwo}%
}{%
 \providecommand\@@startlink[1]{\leavevmode\special{html:<a href="#1">}}%
 \providecommand\@@endlink[0]{\special{html:</a>}}%
}{%
 \providecommand\@@startlink[1]{%
  \leavevmode
  \pdfstartlink
   attr{/Border[0 0 1 ]/H/I/C[0 1 1]}%
   user{/Subtype/Link/A<</Type/Action/S/URI/URI(#1)>>}%
  \relax
 }%
 \providecommand\@@endlink[0]{\pdfendlink}%
}%
\providecommand \url  [0]{\begingroup\@sanitize \@url }%
\providecommand \@url [1]{\endgroup\@href {#1}{\urlprefix}}%
\providecommand \urlprefix [0]{URL }%
\providecommand \Eprint[0]{\href }%
\@ifxundefined \urlstyle {%
  \providecommand \doi [1]{doi:\discretionary{}{}{}#1}%
}{%
  \providecommand \doi [0]{doi:\discretionary{}{}{}\begingroup
  \urlstyle{rm}\Url }%
}%
\providecommand \doibase [0]{http://dx.doi.org/}%
\providecommand \Doi[1]{\href{\doibase#1}}%
\providecommand \bibAnnote [3]{%
  \BibitemShut{#1}%
  \begin{quotation}\noindent
    \textsc{Key:}\ #2\\\textsc{Annotation:}\ #3%
  \end{quotation}%
}%
\providecommand \bibAnnoteFile [2]{%
  \IfFileExists{#2}{\bibAnnote {#1} {#2} {\input{#2}}}{}%
}%
\providecommand \typeout [0]{\immediate \write \m@ne }%
\providecommand \selectlanguage [0]{\@gobble}%
\providecommand \bibinfo [0]{\@secondoftwo}%
\providecommand \bibfield [0]{\@secondoftwo}%
\providecommand \translation [1]{[#1]}%
\providecommand \BibitemOpen[0]{}%
\providecommand \bibitemStop [0]{}%
\providecommand \bibitemNoStop [0]{.\EOS\space}%
\providecommand \EOS [0]{\spacefactor3000\relax}%
\providecommand \BibitemShut [1]{\csname bibitem#1\endcsname}%
\bibitem{Drozdz_1990}%
  \BibitemOpen
  \bibfield{author}{%
  \bibinfo {author} {\bibfnamefont{S.}~\bibnamefont{Dro\.zd\.z}}, \bibinfo
  {author} {\bibfnamefont{S.}~\bibnamefont{Nishizaki}}, \bibinfo {author}
  {\bibfnamefont{J.}~\bibnamefont{Speth}},\ and\ \bibinfo {author}
  {\bibfnamefont{J.}~\bibnamefont{Wambach}},\ }%
  \bibfield{journal}{%
  \bibinfo {journal} {Phys. Rep.}\ }%
  \textbf{\bibinfo {volume} {197}},\ \bibinfo {pages} {1} (\bibinfo {year}
  {1990})%
  \bibAnnoteFile{NoStop}{Drozdz_1990}%
\bibitem{Soloviev_1976}%
  \BibitemOpen
  \bibfield{author}{%
  \bibinfo {author} {\bibfnamefont{V.~G.}\ \bibnamefont{Soloviev}},\ }%
  \emph{\bibinfo {title} {Theory of complex nuclei}}\ (\bibinfo {publisher}
  {Pergamon Press},\ \bibinfo {address} {Oxford},\ \bibinfo {year} {1976})%
  \bibAnnoteFile{NoStop}{Soloviev_1976}%
\bibitem{Krewald_1977}%
  \BibitemOpen
  \bibfield{author}{%
  \bibinfo {author} {\bibfnamefont{J.}~\bibnamefont{Dehesa}}, \bibinfo {author}
  {\bibfnamefont{S.}~\bibnamefont{Krewald}}, \bibinfo {author}
  {\bibfnamefont{J.}~\bibnamefont{Speth}},\ and\ \bibinfo {author}
  {\bibfnamefont{A.}~\bibnamefont{Faessler}},\ }%
  \bibfield{journal}{%
  \bibinfo {journal} {Phys. Rev. C}\ }%
  \textbf{\bibinfo {volume} {15}},\ \bibinfo {pages} {1858} (\bibinfo {year}
  {1977})%
  \bibAnnoteFile{NoStop}{Krewald_1977}%
\bibitem{Tselyaev_1989}%
  \BibitemOpen
  \bibfield{author}{%
  \bibinfo {author} {\bibfnamefont{V.~I.}\ \bibnamefont{Tselyaev}},\ }%
  \bibfield{journal}{%
  \bibinfo {journal} {Yad.Fiz.; Soviet Journal of Nuclear Physics (English
  translation)}\ }%
  \textbf{\bibinfo {volume} {50}},\ \bibinfo {pages} {1252} (\bibinfo {year}
  {1989})%
  \bibAnnoteFile{NoStop}{Tselyaev_1989}%
\bibitem{Kamerdzhiev_1993}%
  \BibitemOpen
  \bibfield{author}{%
  \bibinfo {author} {\bibfnamefont{S.}~\bibnamefont{Kamerdzhiev}}, \bibinfo
  {author} {\bibfnamefont{J.}~\bibnamefont{Speth}}, \bibinfo {author}
  {\bibfnamefont{G.}~\bibnamefont{Tertychny}},\ and\ \bibinfo {author}
  {\bibfnamefont{V.}~\bibnamefont{Tselyaev}},\ }%
  \bibfield{journal}{%
  \Doi{10.1016/0375-9474(93)90315-O}{\bibinfo {journal} {Nucl.Phys.A}}\ }%
  \textbf{\bibinfo {volume} {555}},\ \bibinfo {pages} {90} (\bibinfo {year}
  {1993}),\ ISSN \bibinfo {issn} {0375-9474},\
  \url{http://www.sciencedirect.com/science/article/pii/037594749390315O}%
  \bibAnnoteFile{NoStop}{Kamerdzhiev_1993}%
\bibitem{Kamerdzhiev_2004}%
  \BibitemOpen
  \bibfield{author}{%
  \bibinfo {author} {\bibfnamefont{S.}~\bibnamefont{Kamerdzhiev}}, \bibinfo
  {author} {\bibfnamefont{J.}~\bibnamefont{Speth}},\ and\ \bibinfo {author}
  {\bibfnamefont{G.}~\bibnamefont{Tertychny}},\ }%
  \bibfield{journal}{%
  \Doi{10.1016/j.physrep.2003.11.001}{\bibinfo {journal} {Phys.Rep.}}\ }%
  \textbf{\bibinfo {volume} {393}},\ \bibinfo {pages} {1} (\bibinfo {year}
  {2004}),\ \Eprint{http://arxiv.org/abs/nucl-th/0311058}{arXiv:nucl-th/0311058
  [nucl-th]}%
  \bibAnnoteFile{NoStop}{Kamerdzhiev_2004}%
\bibitem{Lyutorovich_2012}%
  \BibitemOpen
  \bibfield{author}{%
  \bibinfo {author} {\bibfnamefont{N.}~\bibnamefont{Lyutorovich}}, \bibinfo
  {author} {\bibfnamefont{V.~I.}\ \bibnamefont{Tselyaev}}, \bibinfo {author}
  {\bibfnamefont{J.}~\bibnamefont{Speth}}, \bibinfo {author}
  {\bibfnamefont{S.}~\bibnamefont{Krewald}}, \bibinfo {author}
  {\bibfnamefont{F.}~\bibnamefont{Gr{\"{u}}mmer}},\ and\ \bibinfo {author}
  {\bibfnamefont{P.~G.}\ \bibnamefont{Reinhard}},\ }%
  \bibfield{journal}{%
  \Doi{10.1103/PhysRevLett.109.092502}{\bibinfo {journal} {Phys. Rev. Lett.}}\
  }%
  \textbf{\bibinfo {volume} {109}},\ \bibinfo {pages} {092502} (\bibinfo
  {month} {Aug}\ \bibinfo {year} {2012}),\
  \url{http://link.aps.org/doi/10.1103/PhysRevLett.109.092502}%
  \bibAnnoteFile{NoStop}{Lyutorovich_2012}%
\bibitem{Lyutorovich_2015}%
  \BibitemOpen
  \bibfield{author}{%
  \bibinfo {author} {\bibfnamefont{N.}~\bibnamefont{Lyutorovich}}, \bibinfo
  {author} {\bibfnamefont{V.}~\bibnamefont{Tselyaev}}, \bibinfo {author}
  {\bibfnamefont{J.}~\bibnamefont{Speth}}, \bibinfo {author}
  {\bibfnamefont{S.}~\bibnamefont{Krewald}}, \bibinfo {author}
  {\bibfnamefont{F.}~\bibnamefont{Gr\"ummer}},\ and\ \bibinfo {author}
  {\bibfnamefont{P.-G.}\ \bibnamefont{Reinhard}},\ }%
  \bibfield{journal}{%
  \bibinfo {journal} {Phys.Lett.B}\ }%
  \textbf{\bibinfo {volume} {749}},\ \bibinfo {pages} {292} (\bibinfo {year}
  {2015})%
  \bibAnnoteFile{NoStop}{Lyutorovich_2015}%
\bibitem{Ring1}%
  \BibitemOpen
  \bibfield{author}{%
  \bibinfo {author} {\bibfnamefont{D.}~\bibnamefont{Vretenar}}, \bibinfo
  {author} {\bibfnamefont{A.~V.}\ \bibnamefont{Afanasjev}}, \bibinfo {author}
  {\bibfnamefont{G.}~\bibnamefont{Lalazissis}},\ and\ \bibinfo {author}
  {\bibfnamefont{P.}~\bibnamefont{Ring}},\ }%
  \bibfield{journal}{%
  \Doi{10.1016/j.physrep.2004.10.001}{\bibinfo {journal} {Phys. Rep.}}\ }%
  \textbf{\bibinfo {volume} {409}},\ \bibinfo {pages} {101} (\bibinfo {year}
  {2005})%
  \bibAnnoteFile{NoStop}{Ring1}%
\bibitem{Bender_2003}%
  \BibitemOpen
  \bibfield{author}{%
  \bibinfo {author} {\bibfnamefont{M.}~\bibnamefont{Bender}}, \bibinfo {author}
  {\bibfnamefont{P.-H.}\ \bibnamefont{Heenen}},\ and\ \bibinfo {author}
  {\bibfnamefont{P.-G.}\ \bibnamefont{Reinhard}},\ }%
  \bibfield{journal}{%
  \Doi{10.1103/RevModPhys.75.121}{\bibinfo {journal} {Rev.Mod.Phys.}}\ }%
  \textbf{\bibinfo {volume} {75}},\ \bibinfo {pages} {121} (\bibinfo {year}
  {2003})%
  \bibAnnoteFile{NoStop}{Bender_2003}%
\bibitem{Goriely_2002}%
  \BibitemOpen
  \bibfield{author}{%
  \bibinfo {author} {\bibfnamefont{S.}~\bibnamefont{Goriely}}, \bibinfo
  {author} {\bibfnamefont{M.}~\bibnamefont{Samyn}}, \bibinfo {author}
  {\bibfnamefont{P.~H.}\ \bibnamefont{Heenen}}, \bibinfo {author}
  {\bibfnamefont{J.~M.}\ \bibnamefont{Pearson}},\ and\ \bibinfo {author}
  {\bibfnamefont{F.}~\bibnamefont{Tondeur}},\ }%
  \bibfield{journal}{%
  \Doi{10.1103/PhysRevC.66.024326}{\bibinfo {journal} {Phys. Rev. C}}\ }%
  \textbf{\bibinfo {volume} {66}},\ \bibinfo {pages} {024326} (\bibinfo {month}
  {Aug}\ \bibinfo {year} {2002}),\
  \url{http://link.aps.org/doi/10.1103/PhysRevC.66.024326}%
  \bibAnnoteFile{NoStop}{Goriely_2002}%
\bibitem{Nazarevicz1}%
  \BibitemOpen
  \bibfield{author}{%
  \bibinfo {author} {\bibfnamefont{M.}~\bibnamefont{Kortelainen}}, \bibinfo
  {author} {\bibfnamefont{T.}~\bibnamefont{Lesinski}}, \bibinfo {author}
  {\bibfnamefont{J.}~\bibnamefont{Mor\'e}}, \bibinfo {author}
  {\bibfnamefont{W.}~\bibnamefont{Nazarewicz}}, \bibinfo {author}
  {\bibfnamefont{J.}~\bibnamefont{Sarich}}, \bibinfo {author}
  {\bibfnamefont{N.}~\bibnamefont{Schunck}}, \bibinfo {author}
  {\bibfnamefont{M.~V.}\ \bibnamefont{Stoitsov}},\ and\ \bibinfo {author}
  {\bibfnamefont{S.}~\bibnamefont{Wild}},\ }%
  \bibfield{journal}{%
  \Doi{10.1103/PhysRevC.82.024313}{\bibinfo {journal} {Phys. Rev. C}}\ }%
  \textbf{\bibinfo {volume} {82}},\ \bibinfo {pages} {024313} (\bibinfo {year}
  {2010})%
  \bibAnnoteFile{NoStop}{Nazarevicz1}%
\bibitem{Speth_1977}%
  \BibitemOpen
  \bibfield{author}{%
  \bibinfo {author} {\bibfnamefont{J.}~\bibnamefont{Speth}}, \bibinfo {author}
  {\bibfnamefont{E.}~\bibnamefont{Werner}},\ and\ \bibinfo {author}
  {\bibfnamefont{W.}~\bibnamefont{Wild}},\ }%
  \bibfield{journal}{%
  \bibinfo {journal} {Phys. Rep.}\ }%
  \textbf{\bibinfo {volume} {33}},\ \bibinfo {pages} {127} (\bibinfo {year}
  {1977})%
  \bibAnnoteFile{NoStop}{Speth_1977}%
\bibitem{Rei92a}%
  \BibitemOpen
  \bibfield{author}{%
  \bibinfo {author} {\bibfnamefont{P.-G.}\ \bibnamefont{Reinhard}}\ and\
  \bibinfo {author} {\bibfnamefont{Y.}~\bibnamefont{Gambhir}},\ }%
  \bibfield{journal}{%
  \bibinfo {journal} {Ann. Phys. (Leipzig)}\ }%
  \textbf{\bibinfo {volume} {504}},\ \bibinfo {pages} {598} (\bibinfo {year}
  {1992})%
  \bibAnnoteFile{NoStop}{Rei92a}%
\bibitem{Rei92b}%
  \BibitemOpen
  \bibfield{author}{%
  \bibinfo {author} {\bibfnamefont{P.-G.}\ \bibnamefont{Reinhard}},\ }%
  \bibfield{journal}{%
  \bibinfo {journal} {Ann. Phys. (Leipzig)}\ }%
  \textbf{\bibinfo {volume} {504}},\ \bibinfo {pages} {632} (\bibinfo {year}
  {1992})%
  \bibAnnoteFile{NoStop}{Rei92b}%
\bibitem{Jeukenne_1976}%
  \BibitemOpen
  \bibfield{author}{%
  \bibinfo {author} {\bibfnamefont{J.~P.}\ \bibnamefont{Jeukenne}}, \bibinfo
  {author} {\bibfnamefont{A.}~\bibnamefont{Lejeune}},\ and\ \bibinfo {author}
  {\bibfnamefont{C.}~\bibnamefont{Mahaux}},\ }%
  \bibfield{journal}{%
  \Doi{10.1016/0370-1573(76)90017-X}{\bibinfo {journal} {Phys.Rept.}}\ }%
  \textbf{\bibinfo {volume} {25}},\ \bibinfo {pages} {83} (\bibinfo {year}
  {1976})%
  \bibAnnoteFile{NoStop}{Jeukenne_1976}%
\bibitem{Migdal67}%
  \BibitemOpen
  \bibfield{author}{%
  \bibinfo {author} {\bibnamefont{A.B.Migdal}},\ }%
  \emph{\bibinfo {title} {Theory of Finite Fermi Systems and Application to
  Atomic Nuclei}}\ (\bibinfo {publisher} {Wiley},\ \bibinfo {address} {New
  York},\ \bibinfo {year} {1967})%
  \bibAnnoteFile{NoStop}{Migdal67}%
\bibitem{LanLif9}%
  \BibitemOpen
  \bibfield{author}{%
  \bibinfo {author} {\bibfnamefont{L.~D.}\ \bibnamefont{Landau}}, \bibinfo
  {author} {\bibfnamefont{E.~M.}\ \bibnamefont{Lifshitz}},\ and\ \bibinfo
  {author} {\bibfnamefont{L.~P.}\ \bibnamefont{Pitajevski}},\ }%
  \emph{\bibinfo {title} {{C}ourse of {T}heoretical {P}hysics 9 -- {S}tatisical
  {P}hysics}}\ (\bibinfo {publisher} {Pergamon press},\ \bibinfo {address}
  {Oxford},\ \bibinfo {year} {1980})%
  \bibAnnoteFile{NoStop}{LanLif9}%
\bibitem{Sky59a}%
  \BibitemOpen
  \bibfield{author}{%
  \bibinfo {author} {\bibfnamefont{T.~H.~R.}\ \bibnamefont{Skyrme}},\ }%
  \bibfield{journal}{%
  \bibinfo {journal} {Nucl. Phys.}\ }%
  \textbf{\bibinfo {volume} {9}},\ \bibinfo {pages} {615} (\bibinfo {year}
  {1959})%
  \bibAnnoteFile{NoStop}{Sky59a}%
\bibitem{Neg72a}%
  \BibitemOpen
  \bibfield{author}{%
  \bibinfo {author} {\bibfnamefont{J.~W.}\ \bibnamefont{Negele}}\ and\ \bibinfo
  {author} {\bibfnamefont{D.}~\bibnamefont{Vautherin}},\ }%
  \bibfield{journal}{%
  \bibinfo {journal} {Phys. Rev. C}\ }%
  \textbf{\bibinfo {volume} {5}},\ \bibinfo {pages} {1472} (\bibinfo {year}
  {1972})%
  \bibAnnoteFile{NoStop}{Neg72a}%
\bibitem{Brink72}%
  \BibitemOpen
  \bibfield{author}{%
  \bibinfo {author} {\bibfnamefont{D.}~\bibnamefont{Vautherin}}\ and\ \bibinfo
  {author} {\bibfnamefont{D.}~\bibnamefont{Brink}},\ }%
  \bibfield{journal}{%
  \bibinfo {journal} {Phys. Rev.C}\ }%
  \textbf{\bibinfo {volume} {5}},\ \bibinfo {pages} {626} (\bibinfo {year}
  {1972})%
  \bibAnnoteFile{NoStop}{Brink72}%
\bibitem{Frank_2006}%
  \BibitemOpen
  \bibfield{author}{%
  \bibinfo {author} {\bibfnamefont{F.}~\bibnamefont{Gr{\"{u}}mmer}}\ and\
  \bibinfo {author} {\bibfnamefont{J.}~\bibnamefont{Speth}},\ }%
  \bibfield{journal}{%
  \bibinfo {journal} {J.Phys.G:Nucl.Part.Phys.}\ }%
  \textbf{\bibinfo {volume} {32}},\ \bibinfo {pages} {R193} (\bibinfo {year}
  {2006})%
  \bibAnnoteFile{NoStop}{Frank_2006}%
\bibitem{Kamerdzhiev_1997}%
  \BibitemOpen
  \bibfield{author}{%
  \bibinfo {author} {\bibfnamefont{S.~P.}\ \bibnamefont{Kamerdzhiev}}, \bibinfo
  {author} {\bibfnamefont{G.~Y.}\ \bibnamefont{Tertychny}},\ and\ \bibinfo
  {author} {\bibfnamefont{V.~I.}\ \bibnamefont{Tselyaev}},\ }%
  \bibfield{journal}{%
  \bibinfo {journal} {Fiz. Elem. Chastits At. Yadra; Phys. Part. Nucl.}\ }%
  \textbf{\bibinfo {volume} {28}},\ \bibinfo {pages} {333; 134} (\bibinfo
  {year} {1997})%
  \bibAnnoteFile{NoStop}{Kamerdzhiev_1997}%
\bibitem{Tselyaev_2007}%
  \BibitemOpen
  \bibfield{author}{%
  \bibinfo {author} {\bibfnamefont{V.~I.}\ \bibnamefont{Tselyaev}},\ }%
  \bibfield{journal}{%
  \Doi{10.1103/PhysRevC.75.024306}{\bibinfo {journal} {Phys.Rev. C}}\ }%
  \textbf{\bibinfo {volume} {75}},\ \bibinfo {pages} {024306} (\bibinfo {year}
  {2007}),\ \Eprint{http://arxiv.org/abs/nucl-th/0505031}{arXiv:nucl-th/0505031
  [nucl-th]}%
  \bibAnnoteFile{NoStop}{Tselyaev_2007}%
\bibitem{Tselyaev_2013}%
  \BibitemOpen
  \bibfield{author}{%
  \bibinfo {author} {\bibfnamefont{V.~I.}\ \bibnamefont{Tselyaev}},\ }%
  \bibfield{journal}{%
  \bibinfo {journal} {Phys. Rev. C}\ }%
  \textbf{\bibinfo {volume} {88}},\ \bibinfo {pages} {054301} (\bibinfo {year}
  {2013})%
  \bibAnnoteFile{NoStop}{Tselyaev_2013}%
\bibitem{Stone_2007}%
  \BibitemOpen
  \bibfield{author}{%
  \bibinfo {author} {\bibfnamefont{J.~R.}\ \bibnamefont{Stone}}\ and\ \bibinfo
  {author} {\bibfnamefont{P.~.~G.}\ \bibnamefont{Reinhard}},\ }%
  \bibfield{journal}{%
  \bibinfo {journal} {Prog. Part. Nucl. Phys.}\ }%
  \textbf{\bibinfo {volume} {58}},\ \bibinfo {pages} {587} (\bibinfo {year}
  {2007}),\ \bibinfo {note} {http://www.arxiv.org/abs/nucl-th/0607002},\
  \url{http://dx.doi.org/10.1016/j.ppnp.2006.07.001}%
  \bibAnnoteFile{NoStop}{Stone_2007}%
\bibitem{Erler_2011}%
  \BibitemOpen
  \bibfield{author}{%
  \bibinfo {author} {\bibfnamefont{J.}~\bibnamefont{Erler}}, \bibinfo {author}
  {\bibfnamefont{P.}~\bibnamefont{Kl{\"{u}}pfel}},\ and\ \bibinfo {author}
  {\bibfnamefont{P.~G.}\ \bibnamefont{Reinhard}},\ }%
  \bibfield{journal}{%
  \bibinfo {journal} {J. Phys. G}\ }%
  \textbf{\bibinfo {volume} {38}},\ \bibinfo {pages} {033101} (\bibinfo {year}
  {2011}),\ \url{doi:10.1088/0954-3899/38/3/033101}%
  \bibAnnoteFile{NoStop}{Erler_2011}%
\bibitem{Kluepfel_2009}%
  \BibitemOpen
  \bibfield{author}{%
  \bibinfo {author} {\bibfnamefont{P.}~\bibnamefont{Kl{\"{u}}pfel}}, \bibinfo
  {author} {\bibfnamefont{P.~G.}\ \bibnamefont{Reinhard}}, \bibinfo {author}
  {\bibfnamefont{T.~J.}\ \bibnamefont{B{\"{u}}rvenich}},\ and\ \bibinfo
  {author} {\bibfnamefont{J.~A.}\ \bibnamefont{Maruhn}},\ }%
  \bibfield{journal}{%
  \Doi{10.1103/PhysRevC.79.034310}{\bibinfo {journal} {Phys. Rev. C}}\ }%
  \textbf{\bibinfo {volume} {79}},\ \bibinfo {pages} {034310} (\bibinfo {month}
  {Mar}\ \bibinfo {year} {2009}),\
  \url{http://link.aps.org/doi/10.1103/PhysRevC.79.034310}%
  \bibAnnoteFile{NoStop}{Kluepfel_2009}%
\bibitem{Kortelainen_2010}%
  \BibitemOpen
  \bibfield{author}{%
  \bibinfo {author} {\bibfnamefont{M.}~\bibnamefont{Kortelainen}}, \bibinfo
  {author} {\bibfnamefont{T.}~\bibnamefont{Lesinski}}, \bibinfo {author}
  {\bibfnamefont{J.}~\bibnamefont{Mor{\'{e}}}}, \bibinfo {author}
  {\bibfnamefont{W.}~\bibnamefont{Nazarewicz}}, \bibinfo {author}
  {\bibfnamefont{J.}~\bibnamefont{Sarich}}, \emph{et~al.},\ }%
  \bibfield{journal}{%
  \Doi{10.1103/PhysRevC.82.024313}{\bibinfo {journal} {Phys. Rev. C}}\ }%
  \textbf{\bibinfo {volume} {82}},\ \bibinfo {pages} {024313} (\bibinfo {year}
  {2010}),\ \Eprint{http://arxiv.org/abs/1005.5145}{arXiv:1005.5145 [nucl-th]}%
  \bibAnnoteFile{NoStop}{Kortelainen_2010}%
\bibitem{Myers_1977}%
  \BibitemOpen
  \bibfield{author}{%
  \bibinfo {author} {\bibfnamefont{W.~D.}\ \bibnamefont{Myers}},\ }%
  \emph{\bibinfo {title} {Droplet Model of Atomic Nuclei}}\ (\bibinfo
  {publisher} {IFI/Plenum},\ \bibinfo {address} {New York},\ \bibinfo {year}
  {1977})%
  \bibAnnoteFile{NoStop}{Myers_1977}%
\bibitem{Dob14a}%
  \BibitemOpen
  \bibfield{author}{%
  \bibinfo {author} {\bibfnamefont{J.}~\bibnamefont{Dobaczewski}}, \bibinfo
  {author} {\bibfnamefont{W.}~\bibnamefont{Nazarewicz}},\ and\ \bibinfo
  {author} {\bibfnamefont{P.-G.}\ \bibnamefont{Reinhard}},\ }%
  \bibfield{journal}{%
  \bibinfo {journal} {J. Phys. G}\ }%
  \textbf{\bibinfo {volume} {41}},\ \bibinfo {pages} {074001} (\bibinfo {year}
  {2014})%
  \bibAnnoteFile{NoStop}{Dob14a}%
\bibitem{Erl14b}%
  \BibitemOpen
  \bibfield{author}{%
  \bibinfo {author} {\bibfnamefont{J.}~\bibnamefont{Erler}}\ and\ \bibinfo
  {author} {\bibfnamefont{P.-G.}\ \bibnamefont{Reinhard}},\ }%
  \bibfield{journal}{%
  \bibinfo {journal} {J. Phys. G}\ }%
  \textbf{\bibinfo {volume} {42}},\ \bibinfo {pages} {034026} (\bibinfo {year}
  {2014})%
  \bibAnnoteFile{NoStop}{Erl14b}%
\bibitem{Rei15d}%
  \BibitemOpen
  \bibfield{author}{%
  \bibinfo {author} {\bibfnamefont{P.-G.}\ \bibnamefont{Reinhard}},\ }%
  \bibfield{journal}{%
  \bibinfo {journal} {Phys. Scr.}\ }%
  \textbf{\bibinfo {volume} {91}},\ \bibinfo {pages} {023002} (\bibinfo {year}
  {2015}),\ \url{http://dx.doi.org/10.1088/0031-8949/91/2/023002}%
  \bibAnnoteFile{NoStop}{Rei15d}%
\bibitem{Nazarewicz_2013}%
  \BibitemOpen
  \bibfield{author}{%
  \bibinfo {author} {\bibfnamefont{W.}~\bibnamefont{Nazarewicz}}, \bibinfo
  {author} {\bibfnamefont{P.~G.}\ \bibnamefont{Reinhard}}, \bibinfo {author}
  {\bibfnamefont{W.}~\bibnamefont{Satula}},\ and\ \bibinfo {author}
  {\bibfnamefont{D.}~\bibnamefont{Vretenar}},\ }%
  \bibfield{journal}{%
  \bibinfo {journal} {Eur. Phys. J. A}\ }%
  \textbf{\bibinfo {volume} {50}},\ \bibinfo {pages} {20} (\bibinfo {year}
  {2014}),\ \bibinfo {note} {arXiv:1307.5782},\
  \url{http://dx.doi.org/10.1140/epja/i2014-14020-3}%
  \bibAnnoteFile{NoStop}{Nazarewicz_2013}%
\bibitem{NPA928}%
  \BibitemOpen
  \bibfield{author}{%
  \bibinfo {author} {\bibfnamefont{J.}~\bibnamefont{Speth}}, \bibinfo {author}
  {\bibfnamefont{S.}~\bibnamefont{Krewald}}, \bibinfo {author}
  {\bibfnamefont{F.}~\bibnamefont{Gr\"ummer}}, \bibinfo {author}
  {\bibfnamefont{P.~G.}\ \bibnamefont{Reinhard}}, \bibinfo {author}
  {\bibfnamefont{N.}~\bibnamefont{Lyutorovich}},\ and\ \bibinfo {author}
  {\bibfnamefont{V.}~\bibnamefont{Tselyaev}},\ }%
  \bibfield{journal}{%
  \bibinfo {journal} {Nucl. Phys.}\ }%
  \textbf{\bibinfo {volume} {A928}},\ \bibinfo {pages} {17} (\bibinfo {year}
  {2014})%
  \bibAnnoteFile{NoStop}{NPA928}%
\bibitem{Rin80aB}%
  \BibitemOpen
  \bibfield{author}{%
  \bibinfo {author} {\bibfnamefont{P.}~\bibnamefont{Ring}}\ and\ \bibinfo
  {author} {\bibfnamefont{P.}~\bibnamefont{Schuck}},\ }%
  \emph{\bibinfo {title} {The Nuclear Many-Body Problem}}\ (\bibinfo
  {publisher} {Springer--Verl.},\ \bibinfo {address} {New York, Heidelberg,
  Berlin},\ \bibinfo {year} {1980})%
  \bibAnnoteFile{NoStop}{Rin80aB}%
\bibitem{Lyutorovich_2008}%
  \BibitemOpen
  \bibfield{author}{%
  \bibinfo {author} {\bibfnamefont{N.}~\bibnamefont{Lyutorovich}}, \bibinfo
  {author} {\bibfnamefont{J.}~\bibnamefont{Speth}}, \bibinfo {author}
  {\bibfnamefont{A.}~\bibnamefont{Avdeenkov}}, \bibinfo {author}
  {\bibfnamefont{F.}~\bibnamefont{Gr{\"{u}}mmer}}, \bibinfo {author}
  {\bibfnamefont{S.}~\bibnamefont{Kamerdzhiev}}, \bibinfo {author}
  {\bibfnamefont{S.}~\bibnamefont{Krewald}},\ and\ \bibinfo {author}
  {\bibfnamefont{V.}~\bibnamefont{Tselyaev}},\ }%
  \bibfield{journal}{%
  \Doi{10.1140/epja/i2008-10638-x}{\bibinfo {journal} {Eur.Phys.J.}}\ }%
  \textbf{\bibinfo {volume} {A37}},\ \bibinfo {pages} {381} (\bibinfo {year}
  {2008}),\ \Eprint{http://arxiv.org/abs/0806.2813}{arXiv:0806.2813 [nucl-th]}%
  \bibAnnoteFile{NoStop}{Lyutorovich_2008}%
\bibitem{Avdeenkov_2009}%
  \BibitemOpen
  \bibfield{author}{%
  \bibinfo {author} {\bibfnamefont{A.}~\bibnamefont{Avdeenkov}}, \bibinfo
  {author} {\bibfnamefont{F.}~\bibnamefont{Gr{\"{u}}mmer}}, \bibinfo {author}
  {\bibfnamefont{S.}~\bibnamefont{Kamerdzhiev}}, \bibinfo {author}
  {\bibfnamefont{S.}~\bibnamefont{Krewald}}, \bibinfo {author}
  {\bibfnamefont{N.}~\bibnamefont{Lyutorovich}}, \bibinfo {author}
  {\bibfnamefont{V.}~\bibnamefont{Tselyaev}},\ and\ \bibinfo {author}
  {\bibfnamefont{J.}~\bibnamefont{Speth}},\ }%
  \bibfield{journal}{%
  \Doi{10.3103/S1062873809060203}{\bibinfo {journal} {Bulletin of the Russian
  Academy of Sciences: Physics}}\ }%
  \textbf{\bibinfo {volume} {73}},\ \bibinfo {pages} {792} (\bibinfo {year}
  {2009}),\ ISSN \bibinfo {issn} {1062-8738},\
  \url{http://dx.doi.org/10.3103/S1062873809060203}%
  \bibAnnoteFile{NoStop}{Avdeenkov_2009}%
\bibitem{Sil06}%
  \BibitemOpen
  \bibfield{author}{%
  \bibinfo {author} {\bibfnamefont{T.}~\bibnamefont{Sil}}, \bibinfo {author}
  {\bibfnamefont{S.}~\bibnamefont{Shlomo}}, \bibinfo {author}
  {\bibfnamefont{B.~K.}\ \bibnamefont{Agrawal}},\ and\ \bibinfo {author}
  {\bibfnamefont{P.-G.}\ \bibnamefont{Reinhard}},\ }%
  \bibfield{journal}{%
  \bibinfo {journal} {Phys. Rev. C}\ }%
  \textbf{\bibinfo {volume} {73}},\ \bibinfo {pages} {034316} (\bibinfo {year}
  {2006}),\ \bibinfo {note} {http://www.arxiv.org/abs/nucl-th/0601091},\
  \url{http://link.aps.org/doi/10.1103/PhysRevC.73.034316}%
  \bibAnnoteFile{NoStop}{Sil06}%
\bibitem{Belyaev_1995}%
  \BibitemOpen
  \bibfield{author}{%
  \bibinfo {author} {\bibfnamefont{S.~N.}\ \bibnamefont{Belyaev}}, \bibinfo
  {author} {\bibfnamefont{O.~V.}\ \bibnamefont{Vasiliev}}, \bibinfo {author}
  {\bibfnamefont{V.~V.}\ \bibnamefont{Voronov}}, \bibinfo {author}
  {\bibfnamefont{A.~A.}\ \bibnamefont{Nechkin}}, \bibinfo {author}
  {\bibfnamefont{V.~Y.}\ \bibnamefont{Ponomarev}},\ and\ \bibinfo {author}
  {\bibfnamefont{V.~A.}\ \bibnamefont{Semenov}},\ }%
  \bibfield{journal}{%
  \bibinfo {journal} {Phys.Atom.Nucl.}\ }%
  \textbf{\bibinfo {volume} {58}},\ \bibinfo {pages} {1883} (\bibinfo {year}
  {1995})%
  \bibAnnoteFile{NoStop}{Belyaev_1995}%
\bibitem{Youngblood_2004}%
  \BibitemOpen
  \bibfield{author}{%
  \bibinfo {author} {\bibfnamefont{D.~H.}\ \bibnamefont{Youngblood}}, \bibinfo
  {author} {\bibfnamefont{Y.-W.}\ \bibnamefont{Lui}}, \bibinfo {author}
  {\bibfnamefont{H.~L.}\ \bibnamefont{Clark}}, \bibinfo {author}
  {\bibfnamefont{B.}~\bibnamefont{John}}, \bibinfo {author}
  {\bibfnamefont{Y.}~\bibnamefont{Tokimoto}},\ and\ \bibinfo {author}
  {\bibfnamefont{X.}~\bibnamefont{Chen}},\ }%
  \bibfield{journal}{%
  \Doi{10.1103/PhysRevC.69.034315}{\bibinfo {journal} {Phys. Rev. C}}\ }%
  \textbf{\bibinfo {volume} {69}},\ \bibinfo {pages} {034315} (\bibinfo {month}
  {Mar}\ \bibinfo {year} {2004}),\
  \url{http://link.aps.org/doi/10.1103/PhysRevC.69.034315}%
  \bibAnnoteFile{NoStop}{Youngblood_2004}%
\bibitem{Bra85aR}%
  \BibitemOpen
  \bibfield{author}{%
  \bibinfo {author} {\bibfnamefont{M.}~\bibnamefont{Brack}}, \bibinfo {author}
  {\bibfnamefont{C.}~\bibnamefont{Guet}},\ and\ \bibinfo {author}
  {\bibfnamefont{H.-B.}\ \bibnamefont{H{\aa}kansson}},\ }%
  \bibfield{journal}{%
  \bibinfo {journal} {Phys. Rep.}\ }%
  \textbf{\bibinfo {volume} {123}},\ \bibinfo {pages} {275} (\bibinfo {year}
  {1985})%
  \bibAnnoteFile{NoStop}{Bra85aR}%
\bibitem{Erokhova_2003}%
  \BibitemOpen
  \bibfield{author}{%
  \bibinfo {author} {\bibfnamefont{V.~A.}\ \bibnamefont{Erokhova}}, \bibinfo
  {author} {\bibfnamefont{M.~A.}\ \bibnamefont{Elkin}}, \bibinfo {author}
  {\bibfnamefont{A.~V.}\ \bibnamefont{Izotova}}, \bibinfo {author}
  {\bibfnamefont{B.~S.}\ \bibnamefont{Ishkhanov}}, \bibinfo {author}
  {\bibfnamefont{L.~M.}\ \bibnamefont{Kapitonov}}, \bibinfo {author}
  {\bibfnamefont{E.~I.}\ \bibnamefont{Lileeva}},\ and\ \bibinfo {author}
  {\bibfnamefont{E.~V.}\ \bibnamefont{Shirokov}},\ }%
  \bibfield{journal}{%
  \bibinfo {journal} {Izv.Ross.Akad.Nauk.Ser.Fiz..}\ }%
  \textbf{\bibinfo {volume} {67}},\ \bibinfo {pages} {1479} (\bibinfo {year}
  {2003})%
  \bibAnnoteFile{NoStop}{Erokhova_2003}%
\bibitem{Anders_2013}%
  \BibitemOpen
  \bibfield{author}{%
  \bibinfo {author} {\bibfnamefont{M.~R.}\ \bibnamefont{Anders}}, \bibinfo
  {author} {\bibfnamefont{S.}~\bibnamefont{Shlomo}}, \bibinfo {author}
  {\bibfnamefont{T.}~\bibnamefont{Sil}}, \bibinfo {author}
  {\bibfnamefont{D.~H.}\ \bibnamefont{Youngblood}}, \bibinfo {author}
  {\bibfnamefont{Y.-W.}\ \bibnamefont{Lui}},\ and\ \bibinfo {author}
  {\bibnamefont{Krishichayan}},\ }%
  \bibfield{journal}{%
  \Doi{10.1103/PhysRevC.87.024303}{\bibinfo {journal} {Phys. Rev. C}}\ }%
  \textbf{\bibinfo {volume} {87}},\ \bibinfo {pages} {024303} (\bibinfo {month}
  {Feb}\ \bibinfo {year} {2013}),\
  \url{http://link.aps.org/doi/10.1103/PhysRevC.87.024303}%
  \bibAnnoteFile{NoStop}{Anders_2013}%
\bibitem{Toe88a}%
  \BibitemOpen
  \bibfield{author}{%
  \bibinfo {author} {\bibfnamefont{C.}~\bibnamefont{Toepffer}}\ and\ \bibinfo
  {author} {\bibfnamefont{P.-G.}\ \bibnamefont{Reinhard}},\ }%
  \bibfield{journal}{%
  \bibinfo {journal} {Ann. Phys.}\ }%
  \textbf{\bibinfo {volume} {181}},\ \bibinfo {pages} {1} (\bibinfo {year}
  {1988})%
  \bibAnnoteFile{NoStop}{Toe88a}%
\bibitem{Erler_2010}%
  \BibitemOpen
  \bibfield{author}{%
  \bibinfo {author} {\bibfnamefont{J.}~\bibnamefont{Erler}}, \bibinfo {author}
  {\bibfnamefont{P.}~\bibnamefont{Kl{\"{u}}pfel}},\ and\ \bibinfo {author}
  {\bibfnamefont{P.~.~G.}\ \bibnamefont{Reinhard}},\ }%
  \bibfield{journal}{%
  \bibinfo {journal} {J. Phys. G}\ }%
  \textbf{\bibinfo {volume} {37}},\ \bibinfo {pages} {064001} (\bibinfo {year}
  {2010}),\ \bibinfo {note} {http://www.arxiv.org/abs/1002.0027},\
  \url{http://dx.doi.org/10.1088/0954-3899/37/6/064001}%
  \bibAnnoteFile{NoStop}{Erler_2010}%
\bibitem{Engel_1975}%
  \BibitemOpen
  \bibfield{author}{%
  \bibinfo {author} {\bibfnamefont{Y.~M.}\ \bibnamefont{Engel}}, \bibinfo
  {author} {\bibfnamefont{D.~M.}\ \bibnamefont{Brink}}, \bibinfo {author}
  {\bibfnamefont{K.}~\bibnamefont{Goeke}}, \bibinfo {author}
  {\bibfnamefont{S.~J.}\ \bibnamefont{Krieger}},\ and\ \bibinfo {author}
  {\bibfnamefont{D.}~\bibnamefont{Vautherin}},\ }%
  \bibfield{journal}{%
  \bibinfo {journal} {Nucl. Phys. A}\ }%
  \textbf{\bibinfo {volume} {249}},\ \bibinfo {pages} {215} (\bibinfo {year}
  {1975})%
  \bibAnnoteFile{NoStop}{Engel_1975}%
\bibitem{Bog10aR}%
  \BibitemOpen
  \bibfield{author}{%
  \bibinfo {author} {\bibfnamefont{S.~K.}\ \bibnamefont{Bogner}}, \bibinfo
  {author} {\bibfnamefont{R.}~\bibnamefont{Furnstahl}},\ and\ \bibinfo {author}
  {\bibfnamefont{A.}~\bibnamefont{Schwenk}},\ }%
  \bibfield{journal}{%
  \bibinfo {journal} {Prog. Part. Nucl. Phys.}\ }%
  \textbf{\bibinfo {volume} {65}},\ \bibinfo {pages} {94} (\bibinfo {year}
  {2010})%
  \bibAnnoteFile{NoStop}{Bog10aR}%
\bibitem{RF95}%
  \BibitemOpen
  \bibfield{author}{%
  \bibinfo {author} {\bibfnamefont{P.~G.}\ \bibnamefont{Reinhard}}\ and\
  \bibinfo {author} {\bibfnamefont{H.}~\bibnamefont{Flocard}},\ }%
  \bibfield{journal}{%
  \bibinfo {journal} {Nucl. Phys.}\ }%
  \textbf{\bibinfo {volume} {A 584}},\ \bibinfo {pages} {467} (\bibinfo {year}
  {1995})%
  \bibAnnoteFile{NoStop}{RF95}%
\bibitem{Tho61aB}%
  \BibitemOpen
  \bibfield{author}{%
  \bibinfo {author} {\bibfnamefont{D.~J.}\ \bibnamefont{Thouless}},\ }%
  \emph{\bibinfo {title} {The Quantum Mechanics of Many--Body Systems}}\
  (\bibinfo {publisher} {Academic Press},\ \bibinfo {address} {New York},\
  \bibinfo {year} {1961})%
  \bibAnnoteFile{NoStop}{Tho61aB}%
\bibitem{Rei06a}%
  \BibitemOpen
  \bibfield{author}{%
  \bibinfo {author} {\bibfnamefont{P.-G.}\ \bibnamefont{Reinhard}}, \bibinfo
  {author} {\bibfnamefont{M.}~\bibnamefont{Bender}}, \bibinfo {author}
  {\bibfnamefont{W.}~\bibnamefont{Nazarewicz}},\ and\ \bibinfo {author}
  {\bibfnamefont{T.}~\bibnamefont{Vertse}},\ }%
  \bibfield{journal}{%
  \bibinfo {journal} {Phys. Rev. C}\ }%
  \textbf{\bibinfo {volume} {73}},\ \bibinfo {pages} {014309} (\bibinfo {year}
  {2006})%
  \bibAnnoteFile{NoStop}{Rei06a}%
\bibitem{CBHMS98}%
  \BibitemOpen
  \bibfield{author}{%
  \bibinfo {author} {\bibfnamefont{E.}~\bibnamefont{Chabanat}}, \bibinfo
  {author} {\bibfnamefont{P.}~\bibnamefont{Bonnche}}, \bibinfo {author}
  {\bibfnamefont{P.}~\bibnamefont{Haensel}}, \bibinfo {author}
  {\bibfnamefont{J.}~\bibnamefont{Meyer}},\ and\ \bibinfo {author}
  {\bibfnamefont{R.}~\bibnamefont{Schaefer}},\ }%
  \bibfield{journal}{%
  \bibinfo {journal} {Nucl. Phys.}\ }%
  \textbf{\bibinfo {volume} {A635}},\ \bibinfo {pages} {231} (\bibinfo {year}
  {1998})%
  \bibAnnoteFile{NoStop}{CBHMS98}%
\bibitem{Litvinova_2007}%
  \BibitemOpen
  \bibfield{author}{%
  \bibinfo {author} {\bibfnamefont{E.~V.}\ \bibnamefont{Litvinova}}\ and\
  \bibinfo {author} {\bibfnamefont{V.~I.}\ \bibnamefont{Tselyaev}},\ }%
  \bibfield{journal}{%
  \bibinfo {journal} {Phys. Rev. C}\ }%
  \textbf{\bibinfo {volume} {75}},\ \bibinfo {pages} {054318} (\bibinfo {year}
  {2007})%
  \bibAnnoteFile{NoStop}{Litvinova_2007}%
\bibitem{SLKR95}%
  \BibitemOpen
  \bibfield{author}{%
  \bibinfo {author} {\bibfnamefont{M.~M.}\ \bibnamefont{Sharma}}, \bibinfo
  {author} {\bibfnamefont{G.}~\bibnamefont{Lalazissis}}, \bibinfo {author}
  {\bibfnamefont{J.}~\bibnamefont{K\"onig}},\ and\ \bibinfo {author}
  {\bibfnamefont{P.}~\bibnamefont{Ring}},\ }%
  \bibfield{journal}{%
  \bibinfo {journal} {Phys. Rev. Lett.}\ }%
  \textbf{\bibinfo {volume} {74}},\ \bibinfo {pages} {3744} (\bibinfo {year}
  {1995})%
  \bibAnnoteFile{NoStop}{SLKR95}%
\end{thebibliography}%

\end{document}